\documentclass[11p]{article}

\usepackage{fp}
\usepackage[abspage,user,savepos]{zref}
\makeatletter
\AtBeginDocument{%
  \zref@refused{startpoint}%
    \FPmul\p{\zposy{here}}{1}
    \FPdiv\p{\p}{65536}
    \FPadd\pointoffset{\p}{350}
  \hypersetup{%
    pdfstartpage=\zref@extractdefault{startpoint}{abspage}{1},%
    pdfstartview={XYZ 0 \pointoffset 1}
  }%
} \makeatother

\makeatletter
\def\phantomappendix{%
  \Hy@GlobalStepCount\Hy@linkcounter
  \xdef\@currentHref{appendix*.\the\Hy@linkcounter}%
  \Hy@raisedlink{\hyper@anchorstart{\@currentHref}\hyper@anchorend}%
}
\makeatother

\usepackage{amsthm}
\usepackage{enumerate}
\usepackage{dsfont}
\usepackage{latexsym}
\usepackage{amsmath}
\usepackage{amsfonts}
\usepackage{aliascnt}
\usepackage{graphicx}
\usepackage{caption}
\usepackage{subcaption}
\usepackage{color}
\usepackage[margin=1in]{geometry}
\usepackage{natbib}
\usepackage[pdfpagelabels]{hyperref}

\let\DefaultPar\paragraph
\renewcommand{\paragraph}[1]{\DefaultPar{\textbf{#1}.}}

\newcommand{\argmax}{\operatorname{arg\ max}}

\newcommand{\suchthat}{\;\ifnum\currentgrouptype=16 \middle\fi|\;}
\newcommand{\AutoAdjust}[3]{\mathchoice{ \left #1 #2  \right #3}{#1 #2 #3}{#1 #2 #3}{#1 #2 #3} }
\newcommand{\Abs}[1]{{\AutoAdjust{\lvert}{#1}{\rvert}}}

\newcommand{\InBrackets}[1]{\AutoAdjust{[}{#1}{]}}

\newcommand{\RangeMN}[2]{\AutoAdjust{\{}{{#1}\cdots{#2}}{\}}}
\newcommand{\RangeAB}[2]{\AutoAdjust{[}{{#1},{#2}}{]}}
\newcommand{\RangeAb}[2]{\AutoAdjust{[}{{#1},{#2}}{)}}
\newcommand{\RangeaB}[2]{\AutoAdjust{(}{{#1},{#2}}{]}}
\newcommand{\Rangeab}[2]{\AutoAdjust{(}{{#1},{#2}}{)}}
\newcommand{\RangeN}[1]{\InBrackets{#1}}

\newcommand{\Ex}[2][]{\operatorname{\mathbf E}_{#1}\InBrackets{#2}}
\newcommand{\Exn}[2][]{\operatorname{\widehat{\mathbf E}}_{#1}\InBrackets{#2}}
\newcommand{\Prx}[2][]{\operatorname{\mathbf{Pr}}_{#1}\InBrackets{#2}}
\newcommand{\Reals}{\mathbb{R}}

\newcommand{\PosReals}{\Reals_+}
\newcommand{\PosInts}{\mathbb{N}}

\newcommand{\BigO}[1]{{O\AutoAdjust{(}{{#1}}{)}}}

\newcommand{\PhantomPillar}{\rule{0pt}{1em}}

\newcommand{\RV}[1]{{\mathit {#1}}}

\newcommand{\MVec}[1]{{\mathbf #1}}

\newcommand{\RSOPP}{{\RSOP^*}}
\newcommand{\N}{n}

\newcommand{\NN}{\ell}
\newcommand{\Bid}[1]{v_{#1}}
\newcommand{\BidV}{\MVec{v}}
\newcommand{\EQBidV}[1]{{\MVec{q}}^{(#1)}}
\newcommand{\EQBid}[2]{{q}^{(#1)}_{#2}}
\newcommand{\ASet}{{\rm A}}
\newcommand{\A}{{\mathbf A}}
\newcommand{\B}{{\mathbf B}}
\newcommand{\Ev}{{\mathcal E}}
\newcommand{\Av}{{\mathcal A}}
\newcommand{\AvNot}{{\overline{\Av}}}
\newcommand{\Evt}[2]{\Ev_{#1}^{#2}}
\newcommand{\EvtNot}[2]{\overline{\Ev}_{#1}^{#2}}
\newcommand{\EvtJ}[3]{\Ev_{#1,#3}^{#2}}
\newcommand{\EvtDelta}[3]{\Ev_{\RangeaB{#1}{#2}}^{#3}}

\newcommand{\XRV}{\RV{X}}
\newcommand{\Event}{\Ev}
\newcommand{\SRV}{\RV{S}}
\newcommand{\ZRV}{\RV{Z}}
\newcommand{\Rev}{\operatorname{Rev}}%
\newcommand{\RevP}{\operatorname{Rev}^*}%
\newcommand{\Rate}[1]{r_{#1}}
\newcommand{\ARV}{\RV{A}}
\newcommand{\ARVNot}{\overline{\ARV}}
\newcommand{\DiffSeq}{d}
\newcommand{\DiffSeqL}{\ell}
\newcommand{\DiffSeqR}{r}
\newcommand{\Lattice}{{\rm L}}

\newcommand{\Set}{{\rm T}}
\newcommand{\SetU}{{\rm U}}
\newcommand{\SetP}{{\rm T}}
\newcommand{\ND}{d}
\newcommand{\lambdaP}{\lambda'}
\newcommand{\OPT}{\operatorname{OPT}}
\newcommand{\OPTP}{\OPT'}
\newcommand{\OPTPLB}{\underline{\OPT}'}
\newcommand{\OPTPUB}{\overline{\OPT}'}

\newcommand{\ZLB}[2]{{{\rho}_{#1}^{#2}}}
\newcommand{\RevAPLB}[1]{{{r}}^{#1}}
\newcommand{\PriceSetP}[1]{{\overline{\rm A}}^{#1}}
\newcommand{\SP}[2]{s_{#1}^{#2}}
\newcommand{\ZP}[2]{z_{#1}^{#2}}

\newcommand{\BidLB}{\underline{v}}
\newcommand{\BidUB}{\overline{v}}

\newcommand{\NumSlices}{\theta}
\newcommand{\NumSlicesP}{\NumSlices'}
\newcommand{\Const}{c}
\newcommand{\ConstA}{a}

\newcommand{\ResRSOP}{\operatorname{\textsc{RSOPMinExpect}}}

\newcommand{\NumH}{k}

\newcommand{\SetBids}{{\rm I}}
\newcommand{\SetBidsEQ}{{\rm Q}}
\newcommand{\SetBidsEQH}[1]{\SetBidsEQ^{(#1)}}
\newcommand{\Price}{p}

\DeclareMathOperator{\RSOP}{RSOP}%

\newtheorem{theorem}{Theorem}%
\newtheorem{lemma}{Lemma}%
\newtheorem*{lemma*}{Lemma}%

\newtheorem{definition}{Definition}

\newtheorem{proposition}{Proposition}%

\def\more-auths{
\end{tabular}
\begin{tabular}{c}}

\title{On Random Sampling Auctions for Digital Goods
\footnote{A preliminary version appeared in the 10th ACM conference on Electronic Commerce, 2009.}%
\footnote{$ $Id: rsop.tex 55 2013-03-10 23:35:24Z saeed $ $}%
}
\date{}

\author{
Saeed Alaei
    \thanks{Department of Computer Science Cornell University Ithaca, NY 14853.
    \texttt{saeed@cs.cornell.edu}. Supported in part by NSF Award CNS-0720528.}
\and Azarakhsh Malekian
    \thanks{EECS department, Massachusetts Institute of Technology, Cambridge MA 02139. \texttt{malekian@mit.edu}. Supported in part by NSF Award CCF-0728839.}
\and Aravind Srinivasan
    \thanks{Department of Computer Science and Institute for Advanced Computer Studies, University of Maryland, College Park, MD 20742. \texttt{srin@cs.umd.edu}.
    Supported in part by NSF ITR Award CNS-0426683, NSF Award CNS-0626636, and NSF Award CNS 1010789.}
}

\begin{document}
%\conferenceinfo{EC'09,} {July 6--10, 2009, Stanford, California, USA.}%
%\CopyrightYear{2009}%
%\crdata{978-1-60558-458-4/09/07}%

\maketitle

\begin{abstract} In the context of auctions for digital goods, an interesting random sampling auction
has been proposed by \citet*{GHW01}. This auction has been analyzed by \citet*{FFHK05}, who have shown that
it is $15$-competitive in the worst case -- which is substantially better than the previously proven constant
bounds but still far from the conjectured competitive ratio of $4$. In this paper, we prove that the
aforementioned random sampling auction is indeed $4$-competitive for a large class of instances where the
number of bids above (or equal to) the optimal sale price is at least $6$. We also show that it is
$4.68$-competitive for the small class of remaining instances thus leaving a negligible gap between the lower
and upper bound. We employ a mix of probabilistic techniques and dynamic programming to compute these bounds.
\end{abstract}

%\category{F.2.0}{Theory of Computation}{ANALYSIS OF ALGORITHMS AND PROBLEM COMPLEXITY}{General}
%
%\category{G.3}{Mathematics of Computing}{PROBABILITY AND STATISTICS}[Probabilistic algorithms]
%
%\terms{Algorithms, Design, Economics, Theory}
%
%\keywords{Random Sampling, Auction, Mechanism Design}

%\input{rsop_introduction.tex}
%\pagebreak[4]
\section{Introduction}

In recent years, there has been a considerable amount of work in algorithmic mechanism design. Most of this
work can be divided into two categories based on their assumption about prior: (i) Bayesian, and (ii) prior
free. Bayesian mechanism design is based on exploiting the knowledge of the prior to optimize the expected
performance, whereas prior free mechanism design is aimed at optimizing the worst case performance. Random
sampling is perhaps the most popular technique in prior free mechanism design, yet an accurate analysis of
its performance has proven difficult even in the simplest applications.

This paper focuses on analyzing the performance of the random sampling auction proposed by \cite{GHW01},
known as the ``Random Sampling Optimal Price ($\RSOP$)'' auction. The basic problem can be described as
follows. A seller has unlimited supply of a good (e.g., a digital good)~\footnote{If there is a fixed
production cost per copy, the auction can still be used by simply subtracting the production cost from every
bid.} which he is going to sell to unit demand bidders through the following auction: bids are partitioned
into two sets uniformly at random; then the optimal (revenue maximizing) sale price is computed for each set,
and offered as the sale price to the opposite set. The expected revenue of $\RSOP$ is then compared against
the optimal revenue of single price sale of at least two copies.

Most of our analysis is based on the following approach: we develop a lower bound on the performance of
$\RSOP$ that depends on the level of \emph{balancedness} of the partitions, but independent of the bid
values; we then take the expectation of this lower bound over the varying level of balancedness to obtain a
general lower bound on the performance of $\RSOP$. That is in contrast to the previous work based on showing
that a certain level of balancedness is met with a reasonable probability, which inevitably requires a
tradeoff between how strong the balancedness condition is versus how likely it holds.

\paragraph{Related work}
The random sampling optimal price (RSOP) auction has been proposed by \cite{GHW01}, but the problems was
first studied by \cite{GH01}. The revenue of $\RSOP$ has been shown to be close to optimal for many classes
of interesting inputs by \cite{S03}, and \cite{BBHM05}. There has also been a fair amount of work analyzing
the competitive ratio of $\RSOP$. \cite{GH01} showed that $\RSOP$ obtains a constant fraction of the optimal
revenue, and conjectured the constant to be $1/4$; note that the conjecture is tight for an instance with $2$
bidders with distinct bids. A better analysis was proposed by~\cite{FFHK05} which proved the constant to be
at least $1/15$ .

It is important to prove that $\RSOP$ is 4-competitive, because it is a natural and popular mechanism which
is easily implementable and adaptable to various settings (e.g., double auctions~\cite{BV03}, online
limited-supply auctions~\cite{HKP04}, combinatorial auctions~\cite{BBHM05, GH01}, and other setting such
as~\cite{HR08}). Indeed the results of this paper have been used in analysis of other auctions such as the
random sampling based auction of \cite{DH09}  for limited and online supply.

\paragraph{Results}
%To describe our results, we will need the notion of ``winners'' (w.r.t.\ the optimal single-price auction).
%In our definition of $OPT(I)$ where we set $\lambda = \mbox{argmax}_{i \geq 2} ~i \cdot \Bid{i}$, let
%$\lambda$ be the \textbf{largest} index that satisfies this definition. Recall that in the ``offline'' case
%where we know all the $\Bid{i}$ and compute $\lambda$ as this maximizing index, we sell at the single price
%$\Bid{\lambda}$, which is then bought by bidders $1, 2, \ldots, \lambda$ to give an optimal revenue $OPT(I) =
%\lambda \cdot \Bid{\lambda}$ to the auctioneer. Since the number of bidders who get the good in this case is
%$\lambda$, we refer to $\lambda$ as the number of ``winners'' (w.r.t.\ the optimal single-price auction).
%Note that $\lambda$ is determined uniquely by the values $\Bid{1} \geq \Bid{2} \geq \cdots \geq \Bid{N}$.
%
%Many of our results are motivated by the following question: the instance seen above where $n = \lambda = 2$
%and the competitive ratio of RSOP is $4$, seems quite unique. In particular, when selling a digital good, one
%expects the typical number of buyers to be ``large''. Does RSOP do much better than known before, when
%$\lambda$ is large? Our main results are three-fold as follows, and are obtained by an improved probabilistic
%analysis aided by a dynamic programming computation and correlation inequalities:

The following is a summary of our main results.

\begin{enumerate}[I.]
\item
\textbf{Improved lower bounds:} We prove that the ratio of the expected revenue of $\RSOP$ to its benchmark
is:
    \begin{itemize}
        \item at least $1/4.68$ (e.g., \autoref{thm:basic}, and \autoref{thm:xsearch}), improving the previous lower-bound of $1/15$
        due to~\cite{FFHK05};
        \item at least $1/4$, if there are at least 6 bids above (or equal to) the sale price.
        \item at least $1/3.52$, as the number of bids above (or equal to) the sale price approaches infinity.
    \end{itemize}

Our analysis suggests that the worst case performance of $\RSOP$ is attained when there are only two bidders
with distinct bids.

\item
\textbf{Upper bound:} We show that there exist instances where the expected revenue of $\RSOP$ is still less
than $1/2.65$ of its benchmark, even when the number of bids above the optimal sale price approaches
infinity.

\item
\textbf{Combinatorial approach:} We also present a combinatorial lower bound on the performance of $\RSOP$
for a special case when each non-zero bid can take one of only two possible values.

\end{enumerate}

\section{Preliminaries}
\label{sec:prelim}%
We consider auctioning a digital good to $\N$ bidders whose bids are represented by the vector
$\BidV=(\Bid{1},\ldots,\Bid{\N})$ which is, without loss of generality, sorted in decreasing order.

\begin{definition}[$\RSOP$]
\label{def:rsop}%
The \emph{random sampling optimal price} auction partitions the bids into two sets $\A$ and $\B$ uniformly at
random~\footnote{I.e., each bid independently goes to one of $\A$ or $\B$ with probability $\frac{1}{2}$.},
computes the optimal sale price of each set, and offers it as the sale price to the opposite set.
\end{definition}

\begin{definition}[$\OPT$]
The optimal revenue from single price sale to at least two bidders is
\begin{align}
    \OPT &= \max_{j \ge 2} j v_j.
\end{align}
\end{definition}

See \cite{GHKSW06} for motivation of the definition of $\OPT$ and why it requires selling to at least two
bidders.

%
%We consider auctioning digital goods to $N$ bidders with bid values $\Bid{1}, \Bid{2}, \ldots, \Bid{N}$.
%Without loss of generality, we assume $\Bid{1} \geq \Bid{2} \geq \cdots \geq \Bid{N}$. The Random Sampling
%Optimal Price auction partitions the bids into two sets $A$ and $B$ such that each bid $\Bid{i}$
%independently goes to either of $A$ or $B$ with probability $1/2$. We then compute the optimal price of each
%set (among the two sets $A$ and $B$) and offer it to the other set: note that the optimal price of a sequence
%$G = \langle u_1 \geq u_2 \geq \cdots \geq u_k \rangle$ of bids in nondecreasing order, is $u_{\lambda_G}$
%where $\lambda_G = \argmax_{i \geq 1} i u_i$. (Thus, we will use this definition once with $G = A$ when we
%compute the optimal price for $A$ and offer that price to $B$, and will use this definition again with $G =
%B$ when we compute the optimal price for $B$ and offer that price to $A$.) For our input instance $I =
%\Bid{1}, \Bid{2}, \ldots, \Bid{N}$ of bids, we define the optimal revenue of $I$ as $OPT(I) = \lambda
%\Bid{\lambda}$ where $\lambda = \argmax_{i \geq 2} i \Bid{i}$. Note that we force $\lambda \geq 2$ here:
%without this, it can be shown that no incentive compatible mechanism can achieve a constant fraction of the
%optimal revenue in the case where $\Bid{1} \gg \Bid{2}$ . (Note that $\lambda_G$ above is allowed to be one;
%it is only the $\lambda$ that we use in the definition of $OPT(I)$ that is required to be at least two, in
%order to disallow negative results \cite{goldberg01}.)

\paragraph{Assumptions}
We assume $\A$ and $\B$ contain the indices of the bids (as opposed to the actual value of the bids). Without
loss of generality we assume there are infinitely many $0$ bids, i.e., $\Bid{j}=0$ for all $j
> \N$; consequently $(\A,\B)$ is a partitioning of $\PosInts$. The previous assumption allows us to make our
analysis independent of $\N$. Also, without loss of generality we assume $1 \in \B$.~\footnote{Otherwise we
can swap $\A$ and $\B$.}

Throughout most of our analysis we ignore the revenue of $\RSOP$ from bidders in $\A$ because in the
pathological case where $\Bid{1}$ is too large (e.g., $\Bid{1} > \OPT$), the optimal sale price for $\B$ is
equal to $\Bid{1}$ which yields no revenue when offered to $\A$.

\paragraph{Notation}
We adopt the convention of using bold letters for vectors, capital roman letters for sets, capital italic
letters for single-dimensional random variables, capital bold letters for multi-dimensional random variables
(such as sets or vectors), and capital calligraphy letters for events.

We will use $\Ex{\RSOP}$ to denote the expected revenue of $\RSOP$ on an implicit bid vector $\BidV$, where
the expectation is taken over all random partitions $(\A,\B)$; however we sometimes specify an explicit bid
vector by writing $\Ex{\RSOP(\BidV)}$ or $\OPT(\BidV)$.

We use $\lambda$ to denote the index of the optimal sale price which sells to at least two bidders, i.e.,

\begin{align}
    \lambda &\in \argmax_{j \ge 2} j \Bid{j}
\end{align}

%For any $p \in \PosReals$, we denote by $\Rev(\A,p)$ and $\Rev(\B,p)$ the revenue obtained by offering price $p$ to $\A$ and $\B$ respectively.

For every $j \in \PosInts$, we define
\begin{align}
    \SRV_j   &= \Abs{\A\cap \RangeMN{1}{j}}, \\
    \ZRV_j   &= \frac{\Abs{\B\cap \RangeMN{1}{j}}}{\Abs{\A\cap \RangeMN{1}{j}}} = \frac{j-\SRV_j}{\SRV_j}, \label{eq:ZRV_j} \\
    \ZRV     &= \min\left(\{\ZRV_j\}_{j \in \PosInts},1\right) \label{eq:ZRV}.
\end{align}
Note that $\SRV_j$, $\ZRV_j$ and $\ZRV$ are random variables which depend only on how the bids are
partitioned, but not on the actual value of the bids.

For every $\Set \subset \PosInts$ and $\alpha,\alpha' \in \RangeAB{0}{1}$, we define the following events:
\begin{align}
    \Evt{\alpha}{\Set} &= \left\{\PhantomPillar \max_{j \in \Set} \frac{\SRV_j}{j} \le \alpha\right\},\\
    \EvtDelta{\alpha'}{\alpha}{\Set} &= \left\{\PhantomPillar \alpha' < \max_{j \in \Set} \frac{\SRV_j}{j} \le \alpha\right\}=\Evt{\alpha}{\Set} \setminus \Evt{\alpha'}{\Set}
\end{align}
\autoref{fig:event} illustrates an example of $\Evt{\alpha}{\Set}$ and $\EvtDelta{\alpha'}{\alpha}{\Set}$. We
omit $\Set$ if $\Set=\PosInts$, i.e., $\Evt{\alpha}{}=\Evt{\alpha}{\PosInts}$ and
$\EvtDelta{\alpha'}{\alpha}{}=\EvtDelta{\alpha'}{\alpha}{\PosInts}$.

Finally, for any random variable $\XRV$ and event $\Event$, we use $\Exn{\XRV\suchthat \Event}$ to denote the
expectation of $\XRV$ conditioned on event $\Event$ normalized by the probability of $\Event$, i.e.,
\begin{align}
    \Exn{\XRV|\Event} &= \Ex{\XRV \suchthat \Event}\Prx{\Event}.
\end{align}

We will use the following proposition extensively throughout this paper.
\begin{proposition}
\label{prop:exn}%
For any random variable $\XRV$ and any two events $\Event,\Event'$,
\begin{itemize}
\item if $\Event' \subseteq \Event$, then
$\Exn{\XRV \suchthat \Event\setminus\Event'} = \Exn{\XRV \suchthat \Event}-\Exn{\XRV \suchthat \Event'}$;
\item if $\Event \cap \Event' = \emptyset$, then
$\Exn{\XRV \suchthat \Event \cup \Event'} = \Exn{\XRV \suchthat \Event}+\Exn{\XRV \suchthat \Event'}$
\end{itemize}
\end{proposition}

The following lemmas will be useful throughout the rest of this paper.

\begin{lemma}
\label{lem:corr}%
For any $\Set,\Set' \subset \PosInts$ and $\alpha\in\RangeAB{0}{1}$, the two events $\Evt{\alpha}{\Set}$ and
$\Evt{\alpha}{\Set'}$ are positively correlated, i.e., $\Prx{\Evt{\alpha}{\Set}\cap\Evt{\alpha}{\Set'}} \ge
\Prx{\Evt{\alpha}{\Set}}\Prx{\Evt{\alpha}{\Set'}}$ (alternatively
$\Prx{\Evt{\alpha}{\Set}\cap\EvtNot{\alpha}{\Set'}} \le
\Prx{\Evt{\alpha}{\Set}}\Prx{\EvtNot{\alpha}{\Set'}}$).
\end{lemma}
\begin{proof}
The claim follows directly from the FKG inequality and can be found in \autoref{proof:lem:corr}.
\end{proof}

\begin{lemma}
\label{lem:single_tail}%
For any $\alpha \in \Rangeab{0}{1}$ and $j \in \PosInts$,
\begin{align}
    \text{if $\alpha \ge 0.5$, then} \qquad&
        \Prx{\Evt{\alpha}{\{j\}}} \ge 1-\left(\Rate{\alpha}\right)^j,
            &&\text{where} \qquad \Rate{\alpha} = \frac{1}{2\alpha^\alpha (1-\alpha)^{1-\alpha}},\label{eq:rate_alpha}\\
    \text{if $\alpha \le 0.5-1/j$, then} \qquad&
        \Prx{\Evt{\alpha}{\{j\}}} \le \left(\Rate{(\alpha+1/j)}\right)^{j-1}
            && \text{where $\Rate{\alpha}$ is the same as above.}
%    1
\end{align}
\end{lemma}
\begin{proof}
The claim follows from a direct application of Chernoff-Hoeffding bound and can be found in
\autoref{proof:lem:single_tail}.
\end{proof}

%For any $p \in \PosReals$, let $\Rev(\A,p)$ and $\Rev(\B,p)$ denote the revenue obtained by offering price
%$p$ to $\A$ and $\B$ respectively.

\section{The Basic Lower Bound}
\label{sec:basic_bound}%

In this section we prove that RSOP is indeed $4$-competitive for a large class of input instances (i.e., when
$\lambda > 10$). In the next section, we improve this result for $\lambda \le 10$ using a more sophisticated
analysis, but based on the same ideas. The following theorem summarizes the main result of this section.

\begin{theorem}
\label{thm:basic}%
$\Ex{\RSOP} \ge \frac{1}{4}\OPT$ for all $\lambda > 10$. Furthermore, $\Ex{\RSOP} \ge \frac{1}{3.52}\OPT$ for
all $\lambda > 5000$. \autoref{tab:basic_bound} lists the actual lower bounds obtained for various choices of
$\lambda$.
\end{theorem}

The outline of this section is as follows. First, we present a lower bound on $\Ex{\RSOP}$ as a function of
$\lambda$. Recall that expectation is taken over all random partitions $(\A, \B)$ for a fixed set of bids
(and thus a fixed $\lambda$). Our proposed lower bound depends only on $\lambda$ and not the actual value of
the bids. We present a dynamic program for numerically computing the lower bound for any fixed $\lambda$. By
computing the lower bound on $\Ex{\RSOP}$ for all $\lambda \in \RangeMN{11}{5000}$ we confirm that it is
indeed greater than $\frac{1}{4}\OPT$. We then prove a lower bound of $\frac{1}{3.52}\OPT$ on $\Ex{\RSOP}$
for all $\lambda > 5000$.

The following lemma provides a lower bound on $\Ex{\RSOP}$ as a function of $\lambda$.

\begin{lemma}
\label{lem:lbf}%
$\Ex{\RSOP} \ge \Ex{\frac{\SRV_\lambda}{\lambda} \ZRV} \OPT$.
\end{lemma}
\begin{proof}
Let $\Bid{\lambda_{\A}}$ be the optimal price for $\A$ which $\RSOP$ offers to bidders in $\B$; observe that
$\SRV_{\lambda_{\A}} \Bid{\lambda_{\A}} \ge \SRV_j \Bid{j}$ for all $j \in \PosInts$. The revenue of $\RSOP$
is at least the revenue it obtains from $\B$, therefore
\begin{align*}
    \RSOP &\ge (\lambda_{\A}-\SRV_{\lambda_{\A}})\Bid{\lambda_{\A}}
        && \text{because at least $\lambda_{\A}-\SRV_{\lambda_{\A}}$ bids in $\B$ are above or equal to $\Bid{\lambda_{\A}}$} \\
        &= \ZRV_{\lambda_\A} \SRV_{\lambda_{\A}} \Bid{\lambda_{\A}} && \text{by definition of $\ZRV_{\lambda_{\A}}$ in \eqref{eq:ZRV_j}}\\
        &\ge \ZRV \SRV_{\lambda} \Bid{\lambda} && \text{because $\ZRV \le \ZRV_{\lambda_\A}$ and $\SRV_\lambda \Bid{\lambda} \le \SRV_{\lambda_{\A}} \Bid{\lambda_{\A}}$} \\
        &= \ZRV \frac{\SRV_\lambda}{\lambda} \OPT && \text{because $\OPT=\lambda \Bid{\lambda}$. }
\end{align*}
Consequently, $\Ex{\RSOP}\ge \Ex{\frac{\SRV_\lambda}{\lambda} \ZRV} \OPT$ which proves the claim.
\end{proof}

It is crucial that the lower bound provided by the above lemma only depends on $\lambda$ and not on the exact
value of the bids. Recall that $\lambda$ depends only on the value of the bids and not on how the bids are
partitioned.

\subsection{Small $\lambda$}
We start by proving the first part of \autoref{thm:basic}, i.e., that $\Ex{\RSOP} \ge \frac{1}{4}\OPT$ for
all $10 < \lambda \le 5000$.

\begin{figure}
%\centering
\begin{subfigure}[t]{0.5\textwidth}
\centering
\includegraphics[width=\textwidth]{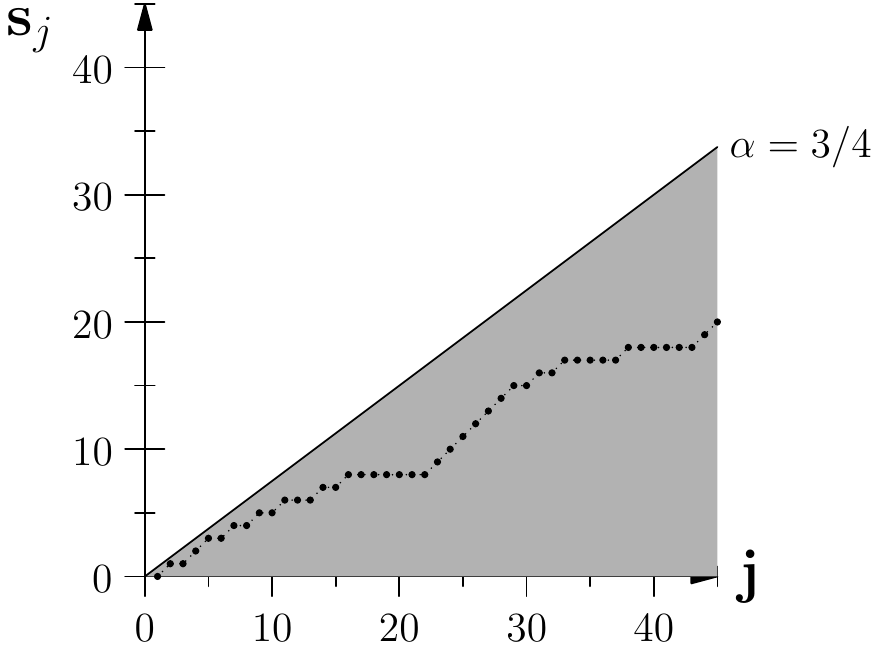}
\caption{$\Evt{\alpha}{}$ denotes the event that the points $(j,\SRV_j)$ lie below the line $y=\alpha x$ (the
gray area). This figure shows the plot of $(j,\SRV_j)$ for an instance of random partitioning in which the
event $\Ev_{\alpha}$ with $\alpha=\frac{3}{4}$ has occurred.}
\end{subfigure}
\hfill
\begin{subfigure}[t]{0.4\textwidth}
\includegraphics[width=\textwidth]{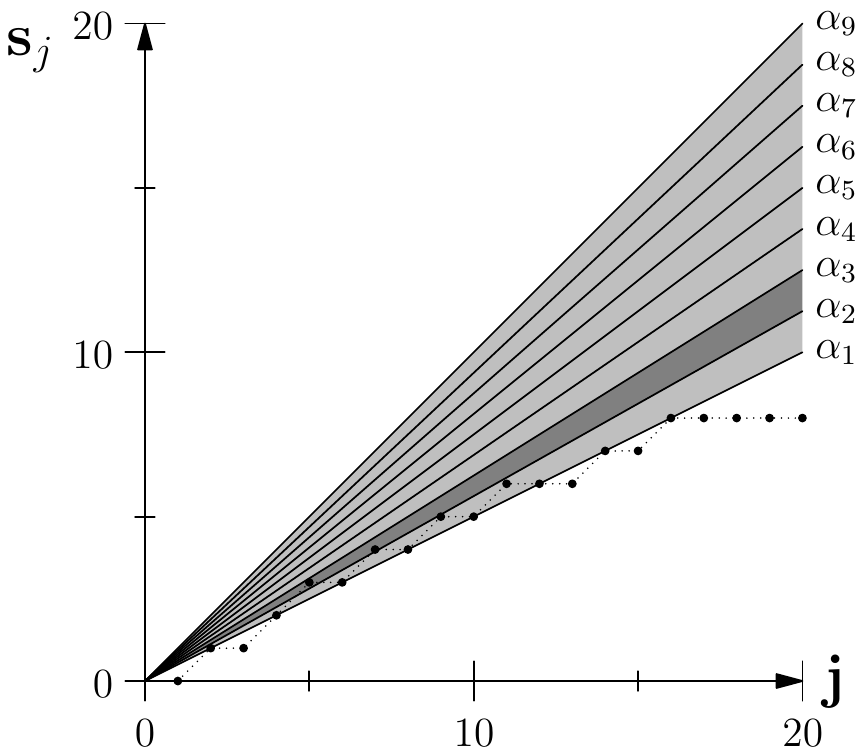}
\caption{$\EvtDelta{\alpha_{i-1}}{\alpha_{i}}{}$ denotes the event that the point $(j,\SRV_j)$ with the
highest ratio of $\frac{\SRV_j}{j}$ lies on or below the line $y = \alpha_{i} x$ and above the line $y =
\alpha_{i-1} x$. This figure shows the plot of $(j,\SRV_j)$ for an instance of random partitioning in which
the event $\EvtDelta{\alpha_2}{\alpha_3}{}$ has occurred (indicated by the dark gray region).}
\end{subfigure}
\caption{\label{fig:event}}
\end{figure}

Recall that $\Ex{\RSOP} \ge \Ex{\frac{\SRV_\lambda}{\lambda} \ZRV} \OPT$ by \autoref{lem:lbf}. Ideally, we
would like to approximate $\Ex{\frac{\SRV_\lambda}{\lambda} \ZRV}$ by
$\Ex{\frac{\SRV_\lambda}{\lambda}}\Ex{\ZRV}$, however $\frac{\SRV_\lambda}{\lambda}$ and $\ZRV$ are
negatively correlated. To work around this obstacle we will decompose $\Ex{\frac{\SRV_\lambda}{\lambda}
\ZRV}$ over a set of small and disjoint events such that, conditioned on each such event, $\ZRV$ can be
approximated closely by a constant. The events are defined as follows. We partition the interval
$\RangeAB{0}{1}$ to small disjoint intervals by picking $m$ points $0.5 < \alpha_1 < \cdots < \alpha_m < 1$.
For each interval $\RangeaB{\alpha_{i-1}}{\alpha_i}$ we consider the event
$\EvtDelta{\alpha_{i-1}}{\alpha_i}{}$. Recall that $\EvtDelta{\alpha_{i-1}}{\alpha_i}{}$ is the event that
$(\max_j \frac{\SRV_j}{j}) \in \RangeaB{\alpha_{i-1}}{\alpha_i}$ (see \autoref{fig:event}). Conditioned on
$\EvtDelta{\alpha_{i-1}}{\alpha_i}{}$, it is easy to see that $\ZRV \in
\RangeAb{\frac{1-\alpha_i}{\alpha_i}}{\frac{1-\alpha_{i-1}}{\alpha_{i-1}}}$, and therefore we can obtain a
good lower bound by substituting $\ZRV$ with $\frac{1-\alpha_i}{\alpha_i}$. Notice that there is no use in
picking $\alpha_i$ from $\RangeAB{0}{0.5}$ because for any $\alpha\in\RangeAB{0}{0.5}$, $\Prx{\Evt{\alpha}{}}
= 0$ and therefore, for any bounded random variable $\XRV$, we get $\Exn{\XRV\suchthat
\Evt{\alpha}{}}=\Ex{\XRV\suchthat \Evt{\alpha}{}}\Prx{\Evt{\alpha}{}} = 0$. Also notice that there is no use
in considering the event $\EvtDelta{\alpha_{m}}{1}{}$ because we can only guarantee a trivial lower bound of
$0$ for $\ZRV$ under $\EvtDelta{\alpha_{m}}{1}{}$.

\begin{lemma}
\label{lem:events}%
Given an increasing sequence $\alpha_1, \ldots, \alpha_m \in \Rangeab{0.5}{1}$, the following inequality
holds for any non-negative random variable $\XRV$.
\begin{align}
    \Ex{\XRV \ZRV} &\ge \sum_{i=1}^{m} \left(\frac{1}{\alpha_i}-\frac{1}{\alpha_{i+1}}\right)\Ex{\XRV \suchthat  \Evt{\alpha_i}{}}.
        \label{eq:events}
\end{align}
Assume $\alpha_{m+1}=1$.
\end{lemma}
\begin{proof}
Let $\alpha_0=0$. We decompose $\Ex{\XRV \ZRV}$ over the set of disjoint events
$\EvtDelta{\alpha_0}{\alpha_{1}}{}, \ldots, \EvtDelta{\alpha_{m-1}}{\alpha_{m}}{}$ as follows.
\begin{align*}
    \Ex{\XRV \ZRV}  &\ge \sum_{i=1}^{m} \Exn{\XRV \ZRV \suchthat \EvtDelta{\alpha_{i-1}}{\alpha_i}{}}
                        && \text{by law of total expectation}\\ %\label{eq:rsop_break}
                    &\ge \sum_{i=1}^{m} \Exn{\XRV \frac{1-\alpha_i}{\alpha_i} \suchthat
                        \EvtDelta{\alpha_{i-1}}{\alpha_i}{}} &&\text{because $\ZRV \ge \frac{1-\alpha_i}{\alpha_i}$ conditioned on $\EvtDelta{\alpha_{i-1}}{\alpha_i}{}$ } \\
                    &= \sum_{i=1}^{m} \frac{1-\alpha_i}{\alpha_i} \left(\Exn{\XRV \suchthat  \Evt{\alpha_{i\vphantom{-1}}}{}}-\Exn{\XRV \suchthat
                        \Evt{\alpha_{i-1}}{}}\right) && \text{by \autoref{prop:exn} given that $\EvtDelta{\alpha_{i-1}}{\alpha_i}{}=\Evt{\alpha_{i\vphantom{-1}}}{}\setminus \Evt{\alpha_{i-1}}{}$}\\
                    &= \sum_{i=1}^{m} \left(\frac{1}{\alpha_i}-\frac{1}{\alpha_{i+1}}\right)\Exn{\XRV \suchthat  \Evt{\alpha_i}{}}
                        && \text{by rearranging the terms}.
     %\label{eq:rsop_alpha_out}
\end{align*}
Note that in the last step we have used the fact that $\Exn{\XRV \suchthat  \Evt{\alpha_0}{}}=0$ (because
$\Prx{\Evt{\alpha_0}{}}=0$).
\end{proof}

The choice of $m$ and $\alpha_1, \ldots, \alpha_m$ in \autoref{lem:events} greatly affects the value of the
lower bound. Generally speaking, increasing $m$ improves the lower bound but at the cost of more computation.

In order to use \autoref{lem:events} effectively, we need to be able to compute
$\Ex{\frac{\SRV_\lambda}{\lambda} \suchthat  \Evt{\alpha_i}{}}$ for each $\alpha_i$. However the events
$\Evt{\alpha_i}{}$ are hard to deal with computationally. The next two lemmas show that
$\Ex{\frac{\SRV_\lambda}{\lambda} \suchthat  \Evt{\alpha_i}{}}$ can be bounded below and thus approximated by
$\Ex{\frac{\SRV_\lambda}{\lambda} \suchthat  \Evt{\alpha_i}{\RangeMN{1}{\NN}}}-\epsilon$ where $\epsilon$
approaches $0$ exponentially fast as a function of $\NN$.

\begin{lemma}
\label{lem:xtail}%
For any random variable $\XRV \in \RangeAB{0}{1}$, any $\alpha \in \RangeaB{0.5}{1}$, and $\NN \in \PosInts$
the following holds:
\begin{align}
    \Exn{\XRV \suchthat \Evt{\alpha}{} \vphantom{\Evt{\alpha}{\RangeMN{1}{\NN}}}} &\ge
        \Exn{\XRV \suchthat \Evt{\alpha}{\RangeMN{1}{\NN}}}-\epsilon & \qquad\text{where}\qquad
        \epsilon &= \Prx{\Evt{\alpha}{\RangeMN{1}{\NN}}} \left(1-\Prx{\Evt{\alpha}{\RangeMN{\NN+1}{\infty}}}\right)
        \label{eq:alpha_event}
\end{align}
\end{lemma}
\begin{proof}%[Proof of \autoref{lem:xtail}]
\label{proof:lem:xtail}%
Observe that $\Evt{\alpha}{}=\Evt{\alpha}{\RangeMN{1}{\NN}} \setminus (\Evt{\alpha}{\RangeMN{1}{\NN}} \cap
\EvtNot{\alpha}{\RangeMN{\NN+1}{\infty}})$ ~\footnote{$\EvtNot{\alpha}{\RangeMN{\NN+1}{\infty}}$ is the
complement of $\Evt{\alpha}{\RangeMN{\NN+1}{\infty}}$.}, therefore
\begin{align*}
    \Exn{\XRV \suchthat \Evt{\alpha}{}\vphantom{\Evt{\alpha}{\RangeMN{1}{\NN}}}}
        &=  \Exn{\XRV \suchthat \Evt{\alpha}{\RangeMN{1}{\NN}}}-\Exn{\XRV \suchthat \Evt{\alpha}{\RangeMN{1}{\NN}} \cap
            \EvtNot{\alpha}{\RangeMN{\NN+1}{\infty}}}
                &&\text{by \autoref{prop:exn}}\\
        &\ge  \Exn{\XRV \suchthat \Evt{\alpha}{\RangeMN{1}{\NN}}}-\Prx{\Evt{\alpha}{\RangeMN{1}{\NN}} \cap
            \EvtNot{\alpha}{\RangeMN{\NN+1}{\infty}}}
                &&\text{because $\XRV \in \RangeAB{0}{1}$} \\
        &\ge  \Exn{\XRV \suchthat \Evt{\alpha}{\RangeMN{1}{\NN}}}-\Prx{\Evt{\alpha}{\RangeMN{1}{\NN}}}
            \left(1-\Prx{\Evt{\alpha}{\RangeMN{\NN+1}{\infty}}}\right)
                &&\text{by \autoref{lem:corr}}
\end{align*}
\end{proof}

The following lemma allows us to compute an upper bound on the $\epsilon$ of the previous lemma.

\begin{lemma}
\label{lem:tail}%
For any $\alpha \in \RangeaB{0.5}{1}$ and any $\NN, \NN' \in \PosInts$ such that $\NN \le \NN'$, the
following holds:
\begin{align}
    \Prx{\Evt{\alpha}{\RangeMN{\NN+1}{\infty}}} & \ge
        \left(1-\frac{\left(\Rate{\alpha}\right)^{\NN'+1}}{1-\Rate{\alpha}}\right) \prod_{j=\NN+1}^{\NN'}
        \left(1-\left(\Rate{\alpha}\right)^j\right) &
        \qquad\text{where $\Rate{\alpha}$ is defined in \eqref{eq:rate_alpha}} \label{eq:tail_bound}
\end{align}
\end{lemma}
\begin{proof}%[Proof of \autoref{lem:tail}]
%Recall that by \autoref{lem:corr}, $\Evt{\alpha}{\Set}$ and $\Evt{\alpha}{\Set'}$ are positively correlated for any $\Set$ and $\Set'$ therefore
\begin{align*}
    \Prx{\Evt{\alpha}{\RangeMN{\NN+1}{\infty}}} = \Prx{\bigcap_{j=\NN+1}^\infty \Evt{\alpha}{\{j\}}}
        &\ge \Prx{\bigcap_{j=\NN'+1}^{\infty} \Evt{\alpha}{\{j\}}} \prod_{j=\NN+1}^{\NN'} \Prx{\Evt{\alpha}{\{j\}}}  &&\text{by \autoref{lem:corr}}\\  %\label{eq:tail_fkg}\\
        &\ge \left(1-\sum_{j=\NN'+1}^{\infty} \Prx{\EvtNot{\alpha}{\{j\}}}\right) \prod_{j=\NN+1}^{\NN'} \Prx{\Evt{\alpha}{\{j\}}} &&\text{by union bound} \\
        &\ge \left(1-\frac{\left(\Rate{\alpha}\right)^{\NN'+1}}{1-\Rate{\alpha}}\right) \prod_{j=\NN+1}^{\NN'}
        \left(1-\left(\Rate{\alpha}\right)^j\right)  &&\text{by \autoref{lem:single_tail}}
\end{align*}
\end{proof}

Observe that in the special case of the above lemma in which $\NN=\NN'$, the right hand side of
\eqref{eq:tail_bound} approaches $1$ exponentially fast as a function of $\NN$ which implies that $\epsilon$
in \eqref{eq:alpha_event} approaches $0$ exponentially fast as a function of $\NN$. Choosing $\NN'>\NN$ only
improves the bound.

The next lemma provides a recurrence relation which can be used to compute the exact value of
$\Ex{\frac{\SRV_\lambda}{\lambda} \suchthat \Evt{\alpha}{\RangeMN{1}{\NN}}}$ and
$\Prx{\Evt{\alpha}{\RangeMN{1}{\NN}}}$ in time $\BigO{\NN^2}$.

\begin{samepage}
\begin{lemma}
\label{lem:dynamic}%
For any $\NN\in \PosInts$ and $\alpha \in \RangeAB{0}{1}$, the exact value of
$\Exn{\frac{\SRV_\lambda}{\lambda} \suchthat \Evt{\alpha}{\RangeMN{1}{\NN}}}$ and
$\Prx{\Evt{\alpha}{\RangeMN{1}{\NN}}}$ can be computed using the following recurrence in which
$\EvtJ{\alpha}{\RangeMN{1}{\NN}}{k}=\Evt{\alpha}{\RangeMN{1}{\NN}}\cap \{\SRV_\NN=k\}$ is the event that
$\Evt{\alpha}{\RangeMN{1}{\NN}}$ happens and $\SRV_\NN=k$.
\begin{align}
    \Prx{\EvtJ{\alpha}{\RangeMN{1}{\NN}}{k}} & =
    \begin{cases}
        \frac{1}{2}\Prx{\EvtJ{\alpha}{\RangeMN{1}{\NN-1}}{k-1}}+\frac{1}{2}\Prx{\EvtJ{\alpha}{\RangeMN{1}{\NN-1}}{k}} \qquad &
            \text{$\NN>1, k\le\alpha \NN$} \\
        1 & \text{$\NN=1, k=0$} \\
        0 & \text{otherwise}
    \end{cases}
    \label{eq:dynamic_p}\\
    \Exn{\frac{\SRV_\lambda}{\lambda} \suchthat \EvtJ{\alpha}{\RangeMN{1}{\NN}}{k}} &=
    \begin{cases}
        \frac{1}{2}\Exn{\frac{\SRV_\lambda}{\lambda} \suchthat \EvtJ{\alpha}{\RangeMN{1}{\NN-1}}{k-1}}+\frac{1}{2}\Exn{\frac{\SRV_\lambda}{\lambda} \suchthat \EvtJ{\alpha}{\RangeMN{1}{\NN-1}}{k}} \qquad &
            \text{$\NN > \lambda, k \le \alpha \NN$} \\
        \frac{k}{\lambda}\Prx{\EvtJ{\alpha}{\RangeMN{1}{\NN}}{k}}  &
            \text{$\NN = \lambda$} \\
        0 & \text{otherwise}
    \end{cases}
    \label{eq:dynamic_e}\\
    \Prx{\Evt{\alpha}{\RangeMN{1}{\NN}}} & = \sum_{\NN=0}^{\NN} \Prx{\EvtJ{\alpha}{\RangeMN{1}{\NN}}{k}} \label{eq:dynamics_p} \\
    \Exn{\frac{\SRV_\lambda}{\lambda}\suchthat \Evt{\alpha}{\RangeMN{1}{\NN}}} & =
        \sum_{k=0}^\NN \Exn{\frac{\SRV_\lambda}{\lambda}\suchthat \EvtJ{\alpha}{\RangeMN{1}{\NN}}{k}} \label{eq:dynamics_e}
\end{align}
\end{lemma}
\end{samepage}
\begin{proof}%[Proof of \autoref{lem:dynamic}]
%\label{proof:lem:dynamic}%

Let $\Av_\NN$ denote the event that $\NN \in \A$. First consider \eqref{eq:dynamic_p}: if $\NN>1$ and
$k\le\alpha \NN$, then $\EvtJ{\alpha}{\RangeMN{1}{\NN}}{k}$ can be decomposed as two disjoint events
$\EvtJ{\alpha}{\RangeMN{1}{\NN-1}}{k-1} \cap \Av_\NN$ and $\EvtJ{\alpha}{\RangeMN{1}{\NN-1}}{k} \cap
\AvNot_\NN$, therefore its probability is the sum of the probabilities of those two event; note that
$\EvtJ{\alpha}{\RangeMN{1}{\NN-1}}{k}$ and $\Av_\NN$ are independent for any $\NN$ and $k$ and
$\Prx{\Av_\NN}=\frac{1}{2}$; furthermore the base of the recursion is $\Prx{\EvtJ{\alpha}{\{1\}}{0}}=1$
because by our assumption $\Av_1=0$ (i.e., the highest bid is always in $\B$). The same argument implies the
correctness of \eqref{eq:dynamic_e} for the case of $\NN>\lambda$. Furthermore,
$\EvtJ{\alpha}{\RangeMN{1}{\lambda}}{k}$ by its definition implies $\SRV_\lambda=k$ which implies the
correctness of \eqref{eq:dynamic_e} for the case of $\NN=\lambda$. Finally \eqref{eq:dynamics_p} and
\eqref{eq:dynamics_e} follow trivially from the law of total probability and the law of total expectation.
\end{proof}

\begin{proof}[Proof of \autoref{thm:basic} for small $\lambda$ (i.e., $10 < \lambda \le 5000$)]
We show how to numerically compute a lower bound on $\Ex{\RSOP}$ for any fixed $\lambda$. Let $m=100$ and
$\alpha_i=0.5+\frac{i}{m+1}$ for each $i \in \RangeN{m}$. Observe that
\begin{align*}
    \Ex{\RSOP}  &\ge \Ex{\frac{\SRV_\lambda}{\lambda} \ZRV} \OPT
                    && \text{by \autoref{lem:lbf}} \\
                &\ge \sum_{i=1}^{m}
                \left(\frac{1}{\alpha_i}-\frac{1}{\alpha_{i+1}}\right)\Ex{\frac{\SRV_\lambda}{\lambda} \suchthat \Evt{\alpha_i}{}} \OPT
                    && \text{by \autoref{lem:events}}
\end{align*}
We then compute a lower bound for each $\Ex{\frac{\SRV_\lambda}{\lambda} \suchthat \Evt{\alpha_i}{}}$ (using
\autoref{lem:xtail}, \autoref{lem:tail}, and \autoref{lem:dynamic} with $\NN=5000$ and $\NN'=100000$), and
substitute them in the last inequality above to obtain a lower bound on $\Ex{\RSOP}$. We have confirmed that
$\Ex{\RSOP} \ge \frac{1}{4}\OPT$ for all $\lambda \in \RangeMN{11}{5000}$ by numerically computing the lower
bound for each choice of $\lambda$ in that range. The computed numerical values of our lower bound are listed
in \autoref{tab:basic_bound} for various choices of $\lambda$.
\end{proof}

\subsection{Large $\lambda$}
We now prove the second part of \autoref{thm:basic}, i.e., $\Ex{\RSOP} \ge \frac{1}{3.52}\OPT$ for all
$\lambda > 5000$.

Recall that $\Ex{\RSOP} \ge \Ex{\frac{\SRV_\lambda}{\lambda} \ZRV} \OPT$ by \autoref{lem:lbf}. Also recall
that $\frac{\SRV_\lambda}{\lambda}$ and $\ZRV$ are negatively correlated, thus
$\Ex{\frac{\SRV_\lambda}{\lambda}}\Ex{\ZRV}$ does not yield a lower bound on
$\Ex{\frac{\SRV_\lambda}{\lambda} \ZRV}$. Nevertheless, the correlation decreases as $\lambda$ increases
which suggests that for sufficiently large $\lambda$ we can separate the two terms. In other words, when
$\lambda$ is large (i.e., $\lambda > 5000$), the two random variables $\frac{\SRV_\lambda}{\lambda}$ and
$\ZRV$ are almost independent and so the expected value of their product is very close to the product of
their expected values. Also for a large $\lambda$ the value of $\frac{\SRV_\lambda}{\lambda}$ is very close
to $\frac{1}{2}$ so $\Ex{\frac{\SRV_\lambda}{\lambda} \ZRV}$ is close to $\frac{1}{2}\Ex{\ZRV}$. We formalize
this argument in the following lemma.

\begin{lemma}
\label{lem:zseparate}%
For any $\alpha \in \RangeAB{0}{1}$:
\begin{align}
    \Ex{\frac{\SRV_\lambda}{\lambda} \ZRV} & \ge
        \alpha \left(\Ex{\vphantom{\Evt{\alpha}{\{\lambda\}}} \ZRV}-\Prx{\Evt{\alpha}{\{\lambda\}}}\right) \label{eq:half_lambda_e2p}
\end{align}
\end{lemma}
\begin{proof}%[Proof of \autoref{lem:zseparate}]
\begin{align*}
    \Ex{\frac{\SRV_\lambda}{\lambda} \ZRV}
        &= \Exn{\frac{\SRV_\lambda}{\lambda} \ZRV \suchthat \EvtNot{\alpha}{\{\lambda\}}}
            +\Exn{\frac{\SRV_\lambda}{\lambda} \ZRV \suchthat \Evt{\alpha}{\{\lambda\}}} \\
        &\ge \alpha \Exn{\ZRV \suchthat \EvtNot{\alpha}{\{\lambda\}}}
                && \text{because $\frac{\SRV_\lambda}{\lambda}>\alpha$ conditioned on $\EvtNot{\alpha}{\{\lambda\}}$} \\
        &= \alpha \left(\Ex{\vphantom{\Evt{\alpha}{\{\lambda\}}}\ZRV}-\Exn{\ZRV \suchthat \Evt{\alpha}{\{\lambda\}}} \right) \\
        &\ge \alpha \left(\Ex{\vphantom{\Evt{\alpha}{\{\lambda\}}}\ZRV}-\Prx{\Evt{\alpha}{\{\lambda\}}}\right)
                && \text{because $\ZRV \le 1$ by its definition, i.e., equation \eqref{eq:ZRV}.}
\end{align*}
\end{proof}

Recall that we can compute an upper bound on $\Prx{\Evt{\alpha}{\{\lambda\}}}$ using
\autoref{lem:single_tail}. Also observe that, for any fixed $\alpha \in \Rangeab{0}{0.5}$,
$\Prx{\Evt{\alpha}{\{\lambda\}}}$ approaches $0$ exponentially fast as a function of $\lambda$ as $\lambda\to
\infty$. The only remaining task is to compute a good lower bound on $\Ex{\ZRV}$.

%
%
%Intuitively, when $\lambda$ is large, in \eqref{eq:half_lambda_e2p}
%the $Pr[\Ev^{[\lambda,\lambda]}_\alpha]$ is very close to $0$. Even
%when $\alpha=\frac{1}{2}-\epsilon$ it roughly gives a lower bound of
%about $\frac{1}{2}E[\Zh]$. Next we show how to compute an upper
%bound on $Pr[\Ev^{[\lambda,\lambda]}_\alpha]$ to support our claim.
%
%\begin{lemma}
%\label{lem:half_lambda_prob}%
%For any $\alpha \in [0, 0.5]$, the following always holds:
%\begin{align}
%    Pr[\Ev^{[\lambda,\lambda]}_\alpha] & \le {C_{\alpha'}}^{\lambda-1} \label{eq:half_lambda_prob} \\
%           \intertext{where} C_{\alpha'} & = \frac{(\frac{1}{\alpha'}-1)^{\alpha'}}{2(1-\alpha')} ,
%           \alpha' = 1-\alpha-\frac{1}{\lambda-1}
%\end{align}
%\end{lemma}
%\begin{proof}
%See \autoref{proof:lem:half_lambda_prob}.
%\end{proof}

\begin{lemma}%[Expected Least Prefix Ratio of $B$ to $A$]
\label{lem:z}%
$\Ex{\ZRV} \ge 0.61$.
\end{lemma}

\begin{proof}
Let $\NN=60000$, $m=100$, and $\alpha_i=0.5+\frac{i}{m+1}$ for each $i \in \RangeN{m}$. By applying
\autoref{lem:events} and plugging $\XRV = 1$ we get
\begin{align*}
    \Ex{\ZRV}   &\ge \sum_{i=1}^{m} \left(\frac{1}{\alpha_i}-\frac{1}{\alpha_{i+1}}\right)\Prx{\Evt{\alpha_i}{}} \\
                &\ge \sum_{i=1}^{m} \left(\frac{1}{\alpha_i}-\frac{1}{\alpha_{i+1}}\right)\Prx{\Evt{\alpha_i}{\RangeMN{1}{\NN}}}\Prx{\Evt{\alpha_i}{\RangeMN{\NN+1}{\infty}}}
                    && \text{by \autoref{lem:corr}}  %\label{eq:z_comp}.
\end{align*}

We then use \autoref{lem:dynamic} to compute $\Prx{\Evt{\alpha_i}{\RangeMN{1}{\NN}}}$, and use
\autoref{lem:tail} with $\NN'=100000$ to compute a lower bound on
$\Prx{\Evt{\alpha_i}{\RangeMN{\NN+1}{\infty}}}$; by substituting the computed numerical values in the above
inequality we get $\Ex{\ZRV} \ge 0.61$.
\end{proof}
It is worth mentioning that by using a similar method we have computed an upper bound of $\Ex{\ZRV} \le 0.63$
which indicates that our analysis is almost tight. \footnote{Note that $1-1/e \simeq 0.6321$ which is
slightly greater that the upper bound of $\Ex{\ZRV} \le 0.63$.}

\begin{proof}[Proof of \autoref{thm:basic} for large $\lambda$ (i.e., $\lambda > 5000$)]
Let $\alpha=0.48$. Then
\begin{align*}
    \Ex{\RSOP}  &\ge \Ex{\frac{\SRV_\lambda}{\lambda} \ZRV} \OPT
                    && \text{by \autoref{lem:lbf}} \\
                &\ge \alpha \left(\Ex{\vphantom{\Evt{\alpha}{\{\lambda\}}} \ZRV}-\Prx{\Evt{\alpha}{\{\lambda\}}}\right) \OPT
                    && \text{by \autoref{lem:zseparate}}. \\
\end{align*}
Using \autoref{lem:single_tail} we get $\Prx{\Evt{\alpha}{\{\lambda\}}} \le 0.0183$ for all $\lambda > 5000$;
furthermore $\Ex{\ZRV} \ge 0.61$ by \autoref{lem:z}; substituting them in the above inequality we get
$\Ex{\RSOP} \ge 0.284$ which is equivalent to a competitive ratio of $3.52$.%
%
%
%on $E[\frac{\SRV_\lambda}{\lambda} \ZRV]$  when $\lambda \ge \bar{\lambda} = 5000$, we used
%\autoref{lem:zseparate} to separate the $\ZRV$ and $\frac{\SRV_\lambda}{\lambda}$ as in
%\eqref{eq:half_lambda_e2p}. Using $E[\ZRV] \ge 0.61$ together with \autoref{lem:half_lambda_prob} and setting
%$\alpha = 0.52$ we get that for any $\lambda > 5000$ , $Pr[\Ev^{[\lambda,\lambda]}_\alpha] \le 0.0183$ and so
%$E[\RSOP] \ge 0.284$ which is equivalent to a competitive ratio of $3.52$ which is better than $4$.
\end{proof}

%??
%Intuitively, $\ZRV$ is a measure of the least ratio of the number of bids in $B$ to the number of bids in $A$
%among any prefix of the bids. A larger $\ZRV$ indicates a more balanced partition. This is an important
%statistic for any random sampling method in general (note that $\ZRV$ only depends on how we partition the
%bids and not the value of the bids).

%\input{rsop_main2.tex}
\section{The Exhaustive Search Lower-Bound}
\label{sec:xsearch_bound}%

In this section we propose an exhaustive search approach which yields an improved lower bound for $\RSOP$ for
small choices of $\lambda$ (i.e., $\lambda \le 10$). The following theorem summarizes the main result of this
section.

\begin{theorem}
\label{thm:xsearch}%
$\Ex{\RSOP} \ge \frac{1}{4.68}\OPT$ for $\lambda \ge 2$ and $\Ex{\RSOP} \ge \frac{1}{4}\OPT$ for $\lambda \ge
6$. \autoref{tab:xsearch} lists the actual lower bounds obtained for various choices of $\lambda$.
\end{theorem}

The basic lower bound of $\Ex{\RSOP}\ge \Ex{\frac{\SRV_\lambda}{\lambda} \ZRV} \OPT$ which we used in
\autoref{sec:basic_bound} does not yield a good enough bound when $\lambda$ is small, mainly because
\begin{enumerate}[(I)]
\item $\frac{\SRV_\lambda}{\lambda}$ and $\ZRV$ are negatively correlated and their correlation is much
stronger when $\lambda$ is small, also
\item the highest bid is always in $\B$, so $\Ex{\frac{\SRV_\lambda}{\lambda}}$ approaches $\frac{1}{4}$ as $\lambda$ goes down to $2$.
\end{enumerate}
Therefore, for $\lambda=2$, $\Ex{\frac{\SRV_\lambda}{\lambda} \ZRV} <
\Ex{\frac{\SRV_\lambda}{\lambda}}\Ex{\ZRV} < 0.25\times 0.63 \approx \frac{1}{6.55}$. In fact, the lower
bounds of \autoref{tab:basic_bound} are quite close to the exact value of $\Ex{\frac{\SRV_\lambda}{\lambda}
\ZRV}$, which suggests for small values of $\lambda$ we need a different approach.

%
%In the rest of this section we explain an \emph{Exhaustive-Search} approach for improving the lower-bound of
%RSOP for the cases where $\lambda$ is small (i.e., $\lambda \le 10$). The basic lower bound of
%$E[\frac{\SRV_\lambda}{\lambda} \ZRV]$ in \autoref{sec:basic_bound} does not work well enough in these cases
%mainly because $\frac{\SRV_\lambda}{\lambda}$ and $\ZRV$ are negatively correlated and their correlation is
%much stronger when $\lambda$ is small. Also because $\Bid{1}$ is always in $B$ and so $\SRV_1$ is always $0$,
%the expected value of $\frac{\SRV_\lambda}{\lambda}$ decreases as $\lambda$ decreases such that for
%$\lambda=2$ we have $\frac{\SRV_\lambda}{\lambda}=\frac{1}{4}$ which is far from $\frac{1}{2}$.

We now provide a high level description of the approach of this section. Without loss of generality we assume
$\OPT=1$. In addition to fixing the index of the optimal price, $\lambda$, we fix the index of the second
optimal price of a higher index, $\lambdaP$, and also fix its corresponding revenue, $\OPTP$, i.e.,
\begin{align}
    \lambdaP    &\in \argmax_{j > \lambda} j\Bid{j} \\
    \OPTP       &=  \max_{j > \lambda} j\Bid{j}
\end{align}
We then try all possible values for $\OPTP$ and $\Bid{1},\ldots,\Bid{\ND}$ (for an appropriate choice of
$\ND$), and apply the techniques from \autoref{sec:basic_bound} to the remaining bids; however instead of
fixing the exact value of each bid and of $\OPTP$ --- which would require checking infinitely many instances
--- we restrict each bid and $\OPTP$ to an interval, i.e., $\OPTP \in \RangeAB{\OPTPLB}{\OPTPUB}$ and
$\Bid{j} \in \RangeAB{\BidLB_j}{\BidUB_j}$ for every $j \in \RangeN{\ND}$. We then try various configurations
of such intervals to cover all possible scenarios. For each configuration we compute a lower bound on
$\Ex{\RSOP}$ for each $\lambda' \le 5000$ as a function of both $\lambda$ and $\lambdaP$, and another lower
bound as a function of only $\lambda$ assuming a reasonably large $\lambdaP$ (e.g., $\lambdaP> 5000$). We
then take the minimum lower bound among all configurations and all $\lambdaP$ to obtain a global lower bound
on $\Ex{\RSOP}$ for each $\lambda \in \RangeMN{2}{10}$; the computed lower bounds are listed in
\autoref{sec:xsearch_bound}.

\begin{lemma}
\label{lem:comb}%
Let $\ResRSOP(\lambda,\{\RangeAB{\BidLB_j}{\BidUB_j}\}_{j \in \RangeN{\ND}}, \RangeAB{\OPTPLB}{\OPTPUB})$
denote the minimum expected revenue of $\Ex{\RSOP}$ over all instances where $\OPTP \in
\RangeAB{\OPTPLB}{\OPTPUB}$, and $\Bid{j} \in \RangeAB{\BidLB_j}{\BidUB_j}$ for all $j \in \RangeN{\ND}$ and
given that $\lambda$ is the index of the optimal price. Then for any $\NumSlices,\NumSlicesP\in \PosInts$,
\begin{align}
    \Ex{\RSOP} &\ge
    \min_{\substack{(i_1,\ldots,i_\ND) \in \RangeN{\NumSlices}^{\ND} \\ i' \in \RangeN{\NumSlicesP}}}
        \ResRSOP\left(\lambda,\left\{\RangeAB{\frac{i_j-1}{\NumSlices}\cdot\frac{1}{j}}{\frac{i_j}{\NumSlices}\cdot\frac{1}{j}}\right\}_{j \in \RangeN{\ND}},
            \RangeAB{\frac{i'-1}{\NumSlicesP}}{\frac{i'}{\NumSlicesP}}\right). \label{eq:comb}
\end{align}
\end{lemma}
\begin{proof}
The claim follows because the minimum is taken over all possible combinations of intervals and that any bid
vector is covered by at least one of the combinations.
\end{proof}

Note that some combinations of intervals in \eqref{eq:comb} might be inconsistent/infeasible; for example it
is infeasible to have both $\RangeAB{\BidLB_2}{\BidUB_2}=\RangeAB{\frac{0}{10}}{\frac{1}{10}}$ and
$\RangeAB{\BidLB_3}{\BidUB_3}=\RangeAB{\frac{4}{15}}{\frac{5}{15}}$ because that would imply
$\Bid{3}>\Bid{2}$; we define $\ResRSOP$ to be $\infty$ if a configuration of intervals is infeasible.

\paragraph{Computing a lower bound on $\ResRSOP$}
In the rest of this section we show how to compute a lower bound on $\Ex{\RSOP}$ given the assumption that
$\OPTP \in \RangeAB{\OPTPLB}{\OPTPUB}$ and $\Bid{j}\in \{\RangeAB{\BidLB_j}{\BidUB_j}\}$ for all $j \in
\RangeN{\ND}$, where these intervals are specified exogenously. The high level idea is to enumerate all
possible partitions of the first $\ND$ bids, define an event for each such partition and decompose
$\Ex{\RSOP}$ over those events, and compute a lower bound conditioned on each such event.

We start with a few definitions. For every $\SetP \subset \RangeN{\ND}$, we define the following event
\begin{align}
    \Av_{\SetP}     &= \left\{\A \cap \RangeMN{1}{\ND} = \SetP \PhantomPillar\right\}
\end{align}
Intuitively, $\Av_{\SetP}$ is the event that, among the first $\ND$ bids, the subset of bids that fall in
$\A$ is exactly $\SetP$. Observe that under the event $\Av_{\SetP}$, both $\SRV_j$ and $\ZRV_j$ are constants
(for every $j \in \RangeN{\ND}$); we will denote those constants respectively by
\begin{align}
    \SP{j}{\SetP}   &=  (\SRV_j \suchthat \Av_{\SetP})
            = \Abs{\RangeMN{1}{j}\cap \SetP} \\
    \ZP{j}{\SetP}   &=  (\ZRV_j \suchthat \Av_{\SetP})
            = \frac{\Abs{\RangeMN{1}{j}\setminus \SetP}}{\Abs{\RangeMN{1}{j}\cap \SetP}}
\end{align}

Our approach is to decompose $\Ex{\RSOP}$ over the set of disjoint events $\{\Av_{\SetP}\}_{\SetP \subseteq
\RangeMN{2}{\ND}}$ and then decompose $\Exn{\RSOP\suchthat \Av_{\SetP}}$ further over the set of disjoint
events $\{\EvtDelta{\alpha_{i-1}}{\alpha_{i}}{\RangeMN{\ND+1}{\infty}}\}_{i \in \RangeN{m}}$ for some choice
of $0.5 < \alpha_1 < \cdots < \alpha_m < 1$ (the second decomposition is similar to
\autoref{sec:basic_bound}); formally,

\begin{align}
    \Ex{\RSOP}  &= \sum_{\SetP \subseteq \RangeMN{2}{\ND}} \sum_{i=1}^{m}  \Exn{\RSOP \suchthat \Av_{\SetP} \cap \EvtDelta{\alpha_{i-1}}{\alpha_i}{\RangeMN{\ND+1}{\infty}}}
                    \label{eq:xdecompose}
\end{align}

Next we show how to compute a lower bound on $\Exn{\RSOP \suchthat \Av_{\SetP} \cap
\EvtDelta{\alpha_{i-1}}{\alpha_i}{\RangeMN{\ND+1}{\infty}}}$ that does not depend on the exact value of the
bids.

\begin{samepage}
\begin{lemma}
\label{lem:xbound}%
For any $\ND \ge 2$, any $\SetP \subseteq \RangeMN{2}{\ND}$, and any $0.5 < \alpha' < \alpha \le 1$,
\begin{align}
    \Exn{\RSOP \suchthat \Av_{\SetP} \cap \EvtDelta{\alpha'}{\alpha}{\RangeMN{\ND+1}{\infty}}}
        &\ge \Exn{\max\left(\RevAPLB{\SetP}, \frac{\SRV_{\lambdaP}}{\lambdaP}\OPTPLB\right) \ZLB{\alpha}{\SetP} \suchthat
            \Av_{\SetP} \cap \EvtDelta{\alpha'}{\alpha}{\RangeMN{\ND+1}{\infty}}}
\end{align}
where $\RevAPLB{\SetP}$ and $\ZLB{\alpha}{\SetP}$ are constants defined as
\begin{align}
    \RevAPLB{\SetP}     &= \max_{j \in \RangeN{\ND}} \SP{j}{\SetP} \BidLB_{j}, \\
    \PriceSetP{\SetP}   &= \left\{j \in \SetP \suchthat \SP{j}{\SetP} \BidUB_{j} \ge \RevAPLB{\SetP} \right\}, \\
    \ZLB{\alpha}{\SetP}  &=
        \begin{cases}
            \min \left\{\ZP{j}{\SetP} \suchthat j \in \PriceSetP{\SetP} \right\}
                &   \text{if $\RevAPLB{\SetP} > \alpha \OPTPUB$}  \\
            \min\left(\left\{\ZP{j}{\SetP} \suchthat j \in \PriceSetP{\SetP}\right\}, \frac{1-\alpha}{\alpha}\right)    &   \text{otherwise}
        \end{cases}.
\end{align}
\end{lemma}
\end{samepage}
\begin{proof}
Let $\Bid{\lambda_{\A}}$ be the optimal price for $\A$ which $\RSOP$ offers to bidders in $\B$; observe that
%$\SRV_{\lambda_{\A}} \Bid{\lambda_{\A}} \ge \SRV_j \Bid{j}$ for all $j \in \PosInts$. The revenue of $\RSOP$
%is at least the revenue it obtains from $\B$, therefore
\begin{align*}
    \RSOP &\ge (\lambda_{\A}-\SRV_{\lambda_{\A}})\Bid{\lambda_{\A}} = \SRV_{\lambda_{\A}} \Bid{\lambda_{\A}} \ZRV_{\lambda_\A}
\end{align*}
Under event $\Av_{\SetP} \cap \EvtDelta{\alpha'}{\alpha}{\RangeMN{\ND+1}{\infty}}$, we show that
$\SRV_{\lambda_{\A}} \Bid{\lambda_{\A}} \ge \max\left(\RevAPLB{\SetP},
\frac{\SRV_{\lambdaP}}{\lambdaP}\OPTPLB\right)$ and $\ZRV_{\lambda_{\A}} \ge \ZLB{\alpha}{\SetP}$, which
combined with the above inequality imply the statement of the lemma.
\begin{itemize}
\item $\SRV_{\lambda_{\A}} \Bid{\lambda_{\A}} \ge \max\left(\RevAPLB{\SetP},
\frac{\SRV_{\lambdaP}}{\lambdaP}\OPTPLB\right)$. Notice that $\SRV_{\lambda_{\A}} \Bid{\lambda_{\A}}$ is the
optimal revenue of $\A$ which must be at least $\RevAPLB{\SetP}$; furthermore, the optimal revenue of $\A$ is
no less than the revenue of selling to $\A$ at price $\Bid{\lambdaP}$ which is at least
$\frac{\SRV_{\lambdaP}}{\lambdaP}\OPTPLB$.

\item $\ZRV_{\lambda_{\A}} \ge \ZLB{\alpha}{\SetP}$.
The inequality follows immediately by considering the following two possibilities:
\begin{enumerate}[(I)]
\item $\lambda_{\A} \le \ND$.
In this case $\lambda_{\A}$ must be in $\PriceSetP{\SetP}$, because for any $j \in
\RangeMN{1}{\ND}\setminus\PriceSetP{\SetP}$, selling to $\A$ at price $\Bid{j}$ generates a revenue which is
less than $\RevAPLB{\SetP}$, therefore $\Bid{j}$ cannot be the optimal price for $\A$.

\item $\lambda_{\A} > \ND$.
First we claim that this case cannot happen if $\RevAPLB{\SetP} > \alpha \OPTPUB$, because otherwise the
revenue of selling to $\A$ at price $\Bid{\lambda_{\A}}$ is less than $\RevAPLB{\SetP}$ which contradicts its
optimality.~\footnote{The revenue of selling to $\A$ at price $\Bid{\lambda_{\A}}$ is $\SRV_{\lambda_{\A}}
\Bid{\lambda_{\A}}$ which is at most $\alpha \lambda_{\A}\Bid{\lambda_{\A}}=\alpha\OPTP$ under event
$\EvtDelta{\alpha'}{\alpha}{\RangeMN{\ND+1}{\infty}}$.}

If indeed $\lambda_{\A} > \ND$, then $\ZRV_{\lambda_{\A}} \ge \frac{1-\alpha}{\alpha}$ under event
$\EvtDelta{\alpha'}{\alpha}{\RangeMN{\ND+1}{\infty}}$.
\end{enumerate}
\end{itemize}
\end{proof}

\begin{lemma}
\label{lem:xevents}%
For any increasing sequence $\alpha_1, \ldots, \alpha_m \in \Rangeab{0.5}{1}$ the following inequality holds
(assume $\alpha_{m+1}=1$).
\begin{align}
    \Ex{\RSOP}  &\ge \sum_{\SetP \subseteq \RangeMN{2}{\ND}} \sum_{i=1}^{m}
                    \left(\ZLB{\alpha_i}{\SetP}-\ZLB{\alpha_{i+1}}{\SetP}\right)  \Exn{\max\left(\RevAPLB{\SetP}, \frac{\SRV_{\lambdaP}}{\lambdaP}\OPTPLB\right)
                    \suchthat \Av_{\SetP} \cap \Evt{\alpha_i}{\RangeMN{\ND+1}{\infty}}}
    \label{eq:xevents}
\end{align}
\end{lemma}
\begin{proof}
The claim follows by applying \autoref{lem:xbound} to equation \eqref{eq:xdecompose}, then decomposing each
event $\EvtDelta{\alpha_{i-1}}{\alpha_i}{\RangeMN{\ND+1}{\infty}}$ as
$\Evt{\alpha_i}{\RangeMN{\ND+1}{\infty}}\setminus \Evt{\alpha_{i-1}}{\RangeMN{\ND+1}{\infty}}$ and applying
\autoref{prop:exn}, and then rearranging the terms.
\end{proof}
%
%Observe that the only random variable on the right hand side of \eqref{eq:xevents} is $\SRV_{\lambdaP}$. So
%we need to compute a numerical lower bound on expressions of the form $\Exn{\max(\Const,
%\frac{\SRV_{\lambdaP}}{\lambdaP}\Const') \suchthat \{\SRV_{\ND}=\ConstA\} \cap
%\Evt{\alpha}{\RangeMN{\ND+1}{\infty}}}$ where $\Const,\Const' \in \RangeAB{0}{1}$ and $\ConstA \in
%\RangeMN{0}{\ND-1}$ are constants~\footnote{It is easy to see that $\Exn{\max(\Const,
%\frac{\SRV_{\lambdaP}}{\lambdaP}\Const') \suchthat \Av_{\SetP} \cap \Evt{\alpha}{\RangeMN{\ND+1}{\infty}}} =
%\Exn{\max(\Const, \frac{\SRV_{\lambdaP}}{\lambdaP}\Const') \suchthat \{\SRV_{\ND}=\Abs{\SetP}\}\cap
%\Evt{\alpha}{\RangeMN{\ND+1}{\infty}}}$.}.

Next we sketch the proof of the main theorem of this section.

\begin{proof}[Proof of \autoref{thm:xsearch}]
We use \autoref{lem:comb} with $\ND=11$, $\NumSlices=3$ and $\NumSlicesP=100$ together with
\autoref{lem:xevents} with $m=100$ and $\alpha_i=0.5+\frac{i}{m+1}$ for each $i \in \RangeN{m}$. To compute
an accurate approximation (lower bound) on each term $\Exn{\max\left(\RevAPLB{\SetP},
\frac{\SRV_{\lambdaP}}{\lambdaP}\OPTPLB\right) \suchthat \Av_{\SetP} \cap
\Evt{\alpha_i}{\RangeMN{\ND+1}{\infty}}}$, we use a combination of dynamic programming and tail bounds
similar to those of \autoref{lem:xtail}, \autoref{lem:tail}, \autoref{lem:dynamic}, \autoref{lem:zseparate},
and \autoref{lem:z} (observe that $\SRV_{\lambdaP}$ is the only random variable in this term). However doing
so naively requires computing a lower bound on as many as $\NumSlices^{\ND-1}\NumSlicesP 2^{\ND-1}m$ such
terms.\footnote{Because there are $\NumSlices^{\ND-1}\NumSlicesP$ possible combinations of intervals in
\eqref{eq:comb} and $2^{\ND-1}$ events of the form $\Av_{\SetP}$ and $m$ events of the form
$\Evt{\alpha_i}{\RangeMN{\ND+1}{\infty}}$.} Instead, we pre-compute $\Exn{\max(\Const,
\SRV_{\lambdaP}\Const') \suchthat \{\SRV_{\ND}=\ConstA\} \cap \Evt{\alpha}{\RangeMN{\ND+1}{\infty}}}$ for all
$\Const,\Const' \in \{\frac{0}{\NumSlicesP},\ldots,\frac{\NumSlicesP}{\NumSlicesP}\}$, all $\ConstA \in
\RangeMN{0}{\ND}$, and all $\alpha \in \{\alpha_1,\ldots,\alpha_m\}$; and then we approximate
$\Exn{\max\left(\RevAPLB{\SetP}, \frac{\SRV_{\lambdaP}}{\lambdaP}\OPTPLB\right) \suchthat \Av_{\SetP} \cap
\Evt{\alpha_i}{\RangeMN{\ND+1}{\infty}}}$ with $\Exn{\max(\Const, \SRV_{\lambdaP}\Const') \suchthat
\{\SRV_{\ND}=\ConstA\} \cap \Evt{\alpha}{\RangeMN{\ND+1}{\infty}}}$ where $\Const,\Const'$ are the result of
rounding $\RevAPLB{\SetP}$ and $\frac{\OPTPLB}{\lambdaP}$ down to the nearest integer multiples of
$\frac{1}{\NumSlicesP}$ respectively and $\ConstA=\Abs{\SetP}$ and $\alpha=\alpha_i$.~\footnote{It is easy to
see that $\Exn{\max(\Const, \frac{\SRV_{\lambdaP}}{\lambdaP}\Const') \suchthat \Av_{\SetP} \cap
\Evt{\alpha}{\RangeMN{\ND+1}{\infty}}} = \Exn{\max(\Const, \frac{\SRV_{\lambdaP}}{\lambdaP}\Const') \suchthat
\{\SRV_{\ND}=\Abs{\SetP}\}\cap \Evt{\alpha}{\RangeMN{\ND+1}{\infty}}}$.} Notice that we only need to
pre-compute $(\NumSlicesP+1)^2\ND m$. \autoref{tab:xsearch} lists the lower bound obtained for each $\lambda
\in \RangeMN{1}{10}$. As a last note, we should mention that we refine each configuration of intervals by
cutting off infeasible regions of each interval prior to any further computation\footnote{For example if
$\BidLB_{j} < \BidLB_{j+1}$, we set $\BidLB_{j} \leftarrow \BidLB_{j+1}$.}.
\end{proof}

\section{An Upper Bound on The Performance of $\RSOP$}
\label{sec:upperbound}%

It has been previously shown that there exist instances of bids for which $\Ex{\RSOP}$ is as low as
$\frac{1}{4}\OPT$ (e.g., \cite{FFHK05,GH01}). However, all such instances have $\lambda = 2$. That raises the
question of whether the performance of $\RSOP$ approaches optimality asymptotically as $\lambda \to \infty$.
In this section, we exhibit a family of instances for which $\Ex{\RSOP}$ is no more than $\frac{1}{2.65}\OPT$
as $\lambda \to \infty$, which proves that the asymptotic competitive ratio of $\RSOP$ is no better than
$2.65$.

\begin{theorem}
\label{thm:upper_bound}%
For any $\lambda \ge 2$ there exists an input instance where there are $\lambda$ bids above or equal to the
optimal sale price and such that $\Ex{\RSOP} < \frac{1}{2.65}\OPT$.
\end{theorem}

Next, we define a family of instances which are used in the proof of the above theorem.

\begin{definition}[Equal Revenue Instance]
\label{def:eq_revenue}%
An instance of bids is called an \emph{equal revenue instance} if choosing any of the bids as the sale price
yields the same revenue. The equal revenue instance with $\N$ non-zero distinct bids is unique (up to
scaling) and given by the bid vector $\EQBidV{\N}=(\EQBid{\N}{1},\EQBid{\N}{2},\ldots)$, where
\begin{align*}
    \EQBid{\N}{j} &=
        \begin{cases}
            \frac{1}{j} & j \le \N \\
            0           & \text{otherwise}
        \end{cases}
\end{align*}
\end{definition}

\begin{proposition}
\label{prop:eq_revenue_rsop}%
For any equal revenue instance, $\RSOP$ offers the worst price to each of the sets $\A$ and $\B$. In other
words, the optimal price of each set generates the least revenue when offered to the opposite set (i.e., less
revenue than offering any of the other non-zero bids as the sale price).
\end{proposition}
\begin{proof}
It follows immediately from the fact that offering any of the bids as the sale price for both sets generates
a total revenue that is equal to $\OPT$. So the price that generates highest revenue for $\A$ also generates
lowest revenue for $\B$ and vice versa.
\end{proof}

\autoref{prop:eq_revenue_rsop} suggests that, for any given $\lambda$, an equal revenue instance might
actually be the worst case instance for $\RSOP$ among all instances with the same $\lambda$; however based on
computer simulation that seems not to be true at least for small values of $\lambda$.

To prove \autoref{thm:upper_bound}, we need to show the expected revenue of $\RSOP$ is no more that
$\frac{1}{2.65}\OPT$ for any equal revenue instance with distinct bids. However a direct analysis of the
performance of $\RSOP$ for all such instances is not easy. Instead we define a modified variant of $\RSOP$
whose performance is easy to analyze, and whose revenue is close to the revenue of $\RSOP$ (e.g.,
asymptotically equal as $\lambda\to \infty$).

\begin{definition}[$\RSOPP$]
\label{def:rsop'}%
The \emph{modified random sampling optimal price auction} behaves exactly the same way as $\RSOP$ (see
\autoref{def:rsop}), except if all of the non-zero bids fall in the same set, it offers them the lowest
non-zero bid as the sale price (instead of $0$) .
\end{definition}

Note that $\RSOPP$ is not a truthful auction, however it is only used to aid the analysis. Next we show that
the revenue of $\RSOP$ is asymptotically equal to the revenue of $\RSOPP$.

\begin{lemma}
\label{lem:rsop_rsop'}%
$\Ex{\RSOP} \le \Ex{\RSOPP} \le \Ex{\RSOP}+\left.(\frac{1}{2})\right.^{\N-1} \OPT$, with the second
inequality being met with equality for equal revenue instances.
\end{lemma}
\begin{proof}
Recall that $\RSOPP$ behaves exactly like $\RSOP$ except when all the $\N$ bids fall in the same set which
happens with probability $\left.(\frac{1}{2})\right.^{\N-1}$, in which case $\RSOPP$ still generates a
revenue of at most $\OPT$ (exactly $\OPT$ if it is an equal revenue instance), while $\RSOP$ generates zero
revenue.
\end{proof}

\begin{lemma}
\label{lem:rsop'_mono}%
$\Ex{\RSOPP(\EQBidV{\N})}$  is a decreasing function of $\N$.
\end{lemma}
\begin{proof}
%Let $\RSOPP(\BidV,\A,\B)$ denote the revenue that $\RSOPP$ generates from the vector of bids $\BidV$ under
%the partition $(\A,\B)$.

Let $\Rev(\BidV,\A,p)$ and $\Rev(\BidV,\B,p)$ denote the revenue obtained by offering price $p$ to bidders
respectively in $\A$ and $\B$ with the vector of bids $\BidV$. Also let $\RevP(\BidV,\A)$ and
$\RevP(\BidV,\B)$ denote the revenue $\RSOPP$ obtains respectively from each of $\A$ and $\B$ under partition
$(\A,\B)$. Observe that $\RevP(\BidV,\A)=0$, because the price that is offered to $\A$ is always $1$. So it
is enough to show that $\Ex{\RevP(\EQBidV{\N},\B)}$ is a decreasing function of $\N$.

Let $\BidV=\EQBidV{\N}$ and $\BidV'=\EQBidV{\N-1}$. We now prove that $\RevP(\BidV,\B) \le \RevP(\BidV',\B)$
which implies the claim of the lemma. Let $\Bid{\lambda_{\A}}$ and $\Bid{\lambda'_{\A}}'$ denote the prices
offered to $\B$ by $\RSOPP$ respectively on $\BidV$ and $\BidV'$, i.e., $\lambda_{\A} \in \argmax_{j \in \A}
\Rev(\BidV, \A, \Bid{j})$, and $\lambda'_{\A} \in \argmax_{j \in \A} \Rev(\BidV', \A, \Bid{j}')$. There are
four possible scenarios:
\begin{enumerate}[(I)]
\item $\N \in \A$ and $\RangeMN{1}{\N-1} \subset \B$.
In this case $\RevP(\BidV,\B) = 1-\frac{1}{\N} < 1 = \RevP(\BidV',\B)$.\footnote{Recall that
$\OPT(\EQBidV{\N})=1$.}

\item $\N \in \A$ and $\RangeMN{1}{\N-1} \not \subset \B$.
In this case either
\begin{enumerate}[(a)]
\item $\lambda_{\A}=\lambda'_{\A} < \N$ and so $\RevP(\BidV,\B) = \RevP(\BidV',\B)$, or
\item $\lambda_{\A} < \lambda'_{\A}=\N$, but that means $\Rev(\BidV,\A,\Bid{\lambda_{\A}}) \ge
\Rev(\BidV',\A,\Bid{\lambda'_{\A}}')$, therefore it must be $\RevP(\BidV,\B)=
\Rev(\BidV,\B,\Bid{\lambda_{\A}}) \le \Rev(\BidV',\B,\Bid{\lambda'_{\A}}')=\RevP(\BidV',\B)$.~\footnote{That
is because both $\BidV$ and $\BidV'$ are equal revenue instances, therefore
$\Rev(\BidV,\B,\Bid{\lambda_{\A}})+\Rev(\BidV,\A,\Bid{\lambda_{\A}})=\Rev(\BidV',\B,\Bid{\lambda'_{\A}}')+\Rev(\BidV',\A,\Bid{\lambda'_{\A}}')=1$.}
\end{enumerate}

\item  $\N \in \B$ and $\RangeMN{1}{\N-1} \subset \B$.
In this case $\RevP(\BidV,\B) = \RevP(\BidV',\B) = 1$.

\item  $\N \in \B$ and $\RangeMN{1}{\N-1} \not \subset \B$.
In this case $\lambda_{\A}=\lambda'_{\A} < \N$ and $\Bid{\lambda_{\A}} > \Bid{\N}$ so $\Bid{\N}$ does not
affect the revenue, therefore $\RevP(\BidV,\B) = \RevP(\BidV',\B)$.

\end{enumerate}
\end{proof}

The following is obtained by direct calculation using a computer.
\begin{proposition}
\label{prop:eqrev_base}%
$\Ex{\RSOPP(\EQBidV{400})}=0.377208 \pm 10^{-6}$.
\end{proposition}

We now prove the main theorem of this section.

\begin{proof}[Proof of \autoref{thm:upper_bound}]
To prove the theorem for any $\lambda$ we exhibit a bid vector $\BidV$ with $\lambda$ bids above the optimal
sale price such that $\Ex{\RSOP(\BidV)} < \frac{1}{2.65} \OPT(\BidV) $. Let $\N=\max(\lambda,400)$, then

\begin{align*}
    \Ex{\RSOP(\EQBidV{\N})} &\le \Ex{\RSOPP(\EQBidV{\N})} &&\text{by \autoref{lem:rsop_rsop'}} \\
                            &\le \Ex{\RSOPP(\EQBidV{400})} &&\text{by \autoref{lem:rsop'_mono}} \\
                            &< 0.377209  &&\text{by \autoref{prop:eqrev_base}} \\
                            &< \frac{1}{2.65} \OPT(\EQBidV{\N})  &&\text{because $\OPT(\EQBidV{\N})=1$} \\
\end{align*}
Observe that the optimal sale price for $\EQBidV{\N}$ is not unique. Let $\BidV$ be the same as $\EQBidV{\N}$
everywhere except $\Bid{\lambda}=\EQBid{\N}{\lambda}+\epsilon$ for a small $\epsilon
\in\Rangeab{0}{\frac{1}{\lambda^2}}$. Observe that $\Bid{\lambda}$ is now the unique optimal sale price for
$\BidV$. It is easy to see that $\lim_{\epsilon\to 0} \Ex{\RSOP(\BidV)}=\Ex{\RSOP(\EQBidV{\N})}$ and
$\lim_{\epsilon\to 0} \OPT(\BidV)=\OPT(\EQBidV{\N})=1$, so for a small enough $\epsilon$, we get
$\Ex{\RSOP(\BidV)} < \frac{1}{2.65} \OPT(\BidV)$ which completes the proof.
\end{proof}

\section{A Combinatorial Lower Bound}

In this section we present a combinatorial approach for obtaining a lower bound on the expected revenue of
$\RSOP$ for equal revenue instance where each non-zero bid is either $1$, or $h$ (for some fixed $h \in
\PosInts$). We hope the ideas we present in the section help develop a more general combinatorial approach in
the future for proving lower bounds on mechanisms based on random sampling.

Observe that in an equal revenue instance where non-zero bids are either $h$ or $1$, if there are $\NumH$
bids of value $h$, there must be $\NumH(h-1)$ bids of vale $1$. Throughout the rest of this section we assume
$h$ is an implicit constant. The following theorem summarizes the main result of this section.

\begin{theorem}
\label{thm:comb}%
For any equal revenue instance where each non-zero bid is either $h$ or $1$,
\begin{align}
    \Ex{\RSOP} \ge \left(\frac{1}{2}+\frac{1}{2h}-\frac{1}{2^{\NumH h-1}}\right)\OPT
\end{align}
where $\NumH$ is the number of bids of value $h$.
\end{theorem}
Observe that in the above theorem the worst case of the lower bound is when $\NumH=1$ and $h=2$ for which the
lower bound becomes $\OPT/4$. Notice that the lower bound approaches $\OPT/2$ quickly as either $\NumH$ or
$h$ increases.

%
%\begin{definition}
%For any $\NumH,\NumL \in \PosIntsZ$, $\CuriousSet{\NumH}{\NumL}$ denotes a multi set consisting of $\NumH$
%bids of value $h$ and $\NumL$ bids of value $1$.
%\end{definition}

\begin{definition}
$\SetBidsEQH{\NumH}$ denotes the multi-set of bid corresponding to an equal revenue instance with $\NumH$
bids of value $h$ and $\NumH(h-1)$ bids of value $1$.
\end{definition}

For the rest of this section we assume that $\A$ and $\B$ are multi-sets containing the actual bids in each
side of the partition, as opposed to the previous sections where we assumed $\A$ and $\B$ contained the
indices of those bids. Furthermore, for any multi-set of bids such as $\SetBids$, we use the notation
$\Ex{\RSOP(\SetBids)}$, $\Ex{\RSOPP(\SetBids)}$ and $\OPT(\SetBids)$ to denote the respective quantity being
computed on bids explicitly specified by $\SetBids$. We also make no assumption about which of $\A$ or $\B$
gets the highest bid, unless explicitly stated otherwise.

We start by proving a lower bound on the expected revenue of $\RSOPP$ (see \autoref{def:rsop'}) on equal
revenue instances where each non-zero bid is either $h$ or $1$. We then extend the lower bound to $\RSOP$.
Recall that $\RSOPP$ behaves exactly the same way as $\RSOP$, except if all non-zero bids fall in the same
set, $\RSOPP$ offers them the lowest non-zero bid as the sale price (instead of $0$).

\begin{lemma}
\label{lem:comb'}%
For any $\NumH \in \PosInts$,
\begin{align}
    \Ex{\RSOPP(\SetBidsEQH{\NumH})} \ge \frac{\NumH (h+1)}{2} = \frac{1}{2}\OPT(\SetBidsEQH{\NumH})+\frac{k}{2}.
\end{align}
\end{lemma}
\begin{proof}
We prove the claim by induction on $\NumH$.
%For every $\NumH \in \PosInts$, let $\SetBidsEQH{\NumH}$ be a multi-set of bids corresponding to an equal
%revenue instance with $\NumH$ bids of value $h$ and $\NumH(h-1)$ bids of value $1$. We show that
%$\Ex{\RSOPP(\SetBidsEQH{\NumH})} \ge \OPT(\SetBidsEQH{\NumH}) = \frac{\NumH h}{2}$ by induction on $\NumH$.

We first prove the base case which is $\NumH=1$. The single bid of value $h$ is the highest bid. Without loss
of generality assume that the $h$ bid is in $\B$. Observe that the optimal price of $\B$ is $h$ which is also
the price offered to $\A$, so no revenue is obtained from $\A$. Furthermore the optimal price of $\A$ is $1$
which is also the price offered to $\B$. Each bid of value $1$ falls in $\B$ with probability $1/2$, so
$\Ex{\RSOPP(\SetBidsEQH{1})}=1+\frac{h-1}{2}$ which proves the base of the induction.

We now prove the induction step. For any two multi-sets of bids such as $\SetP$ and $\SetU$, let
$\RevP(\SetP,\SetU)$ denote the revenue obtained from $\SetP$ by computing the optimal sale price for $\SetU$
(let the optimal price be $1$ if $\SetU=\emptyset$) and offering that price to $\SetP$; also let
$\Rev(\SetP,\Price)$ denote the revenue obtained by offering price $\Price$ to $\SetP$. Let $(\A,\B)$ be a
random partition of $\SetBidsEQH{1}$, and let $(\A',\B')$ be a random partition of $\SetBidsEQH{\NumH-1}$.
Observe that $(\A\cup\A',\B\cup\B')$ is a random partition of $\SetBidsEQH{\NumH}$. The induction step
follows from the following inequities.
\begin{align*}
    \Ex{\RSOPP(\SetBidsEQH{\NumH})} &= \Ex{\RevP(\A\cup\A',\B\cup\B')+\RevP(\B\cup\B',\A\cup\A')} \\
                                    &\ge \Ex{\RevP(\A,\B)+\RevP(\A',\B')+\RevP(\B,\A)+\RevP(\B',\A')}
                                        &&\text{to be proven}\\
                                    &= \Ex{\RSOPP(\SetBidsEQH{1})}+\Ex{\RSOPP(\SetBidsEQH{\NumH-1})} \\
                                    &\ge \frac{h+1}{2}+\frac{(\NumH-1) (h+1)}{2}>\frac{\NumH (h+1)}{2}
                                        && \text{by the induction hypothesis}
\end{align*}
We shall prove $\RevP(\B\cup\B',\A\cup\A')>\RevP(\B,\A)+\RevP(\B',\A')$ and by symmetry we can argue
$\RevP(\A\cup\A',\B\cup\B')>\RevP(\A,\B)+\RevP(\A',\B')$ which completes the proof. Let $\Price$, $\Price'$
and $\Price''$  denote the optimal price of $\A$, $\A'$ and $\A \cup \A'$ respectively as computed by
$\RSOPP$ (i.e., the optimal price for an empty set would be $1$). We argue that
\begin{align*}
    \RevP(\B\cup\B',\A\cup\A')  &= \Rev(\B,\Price'')+\Rev(\B',\Price'') \\
                                &\ge \Rev(\B,\Price)+\Rev(\B',\Price') && \text{explained below} \\
                                &= \RevP(\B,\A)+\RevP(\B',A').
\end{align*}
Observe that both $\A\cup\B$ and $\A'\cup\B'$ are equal revenue instances and by
\autoref{prop:eq_revenue_rsop} in any equal revenue instance the price that is optimal for one side generates
the least revenue for the opposite side so $\Rev(\B,\Price'')\ge \Rev(\B,\Price)$ and $\Rev(\B',\Price'')\ge
\Rev(\B',\Price')$.

%\About{Arbitrary instances}  We now prove the lemma for an arbitrary instance represented by a multi-set of
%bids $\SetBids$. Let $\NumH$ be the largest integer such that $\SetBidsEQH{\NumH}\subset \SetBids$. Let
%$\SetBids'=\SetBids\setminus \SetBidsEQH{\NumH}$. Let $\NumH'$ and $\NumL'$ denote respectively the number of
%bids of value $h$ and the number of bids of value $1$ in $\SetBids'$. Let $(\A,\B)$ be a random partition of
%$\SetBidsEQH{\NumH}$ and $(\A',\B')$ be a random partition of the bids in
\end{proof}

\begin{proof}[Proof of \autoref{thm:comb}]
Recall that the only situation where $\RSOPP$ and $\RSOP$ behave differently is when either $\A$ or $\B$ is
empty which happens with probability $1/2^{\NumH h-1}$, therefore

\begin{align*}
    \Ex{\RSOP(\SetBidsEQH{\NumH})}
            &=  \Ex{\RSOPP(\SetBidsEQH{\NumH})}-\frac{1}{2^{\NumH h-1}} \OPT \\
            &\ge \frac{\NumH (h+1)}{2}-\frac{1}{2^{\NumH h-1}} \OPT && \text{by \autoref{lem:comb'}} \\
            &= \left(\frac{1}{2}+\frac{1}{2h}-\frac{1}{2^{\NumH h-1}}\right) \OPT && \text{because $\OPT=\NumH h$}
\end{align*}
That completes the proof.
\end{proof}

\section{Acknowledgment}
We would like to thank Jason Hartline for several valuable
discussions. We also thank the anonymous referees for their helpful
and detailed comments.

\vspace{\stretch{1}}

\bibliography{rsop}

\vspace{\stretch{1}}

\pagebreak[4]
\appendix

\section{Results}

\begin{table}[h]
\scriptsize

\centering
\begin{tabular}{|l|l|l|}
\hline%
$\lambda$ & $\Ex{\RSOP}/\OPT$ & Competitive-Ratio \\ \hline \hline
    2 & 0.125148 & 7.99 \\ \hline
    3 & 0.166930 & 5.99 \\ \hline
    4 & 0.192439 & 5.20 \\ \hline
    5 & 0.209222 & 4.78 \\ \hline
    6 & 0.221407 & 4.52 \\ \hline
    7 & 0.230605 & 4.34 \\ \hline
    8 & 0.237862 & 4.20 \\ \hline
    9 & 0.243764 & 4.10 \\ \hline
   10 & 0.248647 & 4.02 \\ \hline
   11 & 0.252774 & 3.96 \\ \hline
   15 & 0.264398 & 3.78 \\ \hline
   20 & 0.273005 & 3.66 \\ \hline
   30 & 0.282297 & 3.54 \\ \hline
   50 & 0.290384 & 3.44 \\ \hline
  100 & 0.296993 & 3.37  \\ \hline
  200 & 0.300549 & 3.33  \\ \hline
  300 & 0.301784 & 3.31  \\ \hline
  500 & 0.302792 & 3.30  \\ \hline
 1000 & 0.303560 & 3.29  \\ \hline
 1500 & 0.303818 & 3.29  \\ \hline
 2000 & 0.303949 & 3.29  \\ \hline
\end{tabular}

\caption{Computed numerical values for the basic lower-bound of $\Ex{\RSOP}\ge
\Ex{\frac{\SRV_\lambda}{\lambda}\ZRV}\OPT$. \label{tab:basic_bound}}
\end{table}

\begin{table}[h]
\scriptsize \centering
\begin{tabular}{|l|l|l|}
\hline%
    $\lambda$ & $\Ex{\RSOP}/\OPT$ & Competitive-Ratio \\ \hline \hline
   2 & 0.2138 & 4.68 \\ \hline
   3 & 0.2178 & 4.59 \\ \hline
   4 & 0.238  & 4.20 \\ \hline
   5 & 0.243  & 4.11 \\ \hline
   6 & 0.2503 & 3.99 \\ \hline
   7 & 0.2545 & 3.93 \\ \hline
   8 & 0.2602 & 3.84 \\ \hline
   9 & 0.2627 & 3.81 \\ \hline
  10 & 0.2669 & 3.75 \\ \hline
\end{tabular}
\caption{Computed numerical values for the exhaustive-search lower-bound\label{tab:xsearch}}
\end{table}

\section{Proofs}
\label{sec:proofs}%

\begin{theorem}[Chernoff-Hoeffding\citeyear{hoeffding}]
\label{thm:hoeffding}%
For (i.i.d.) random variables $\XRV_1, \XRV_2, \ldots, \XRV_\NN \in \{0,1\}$ with $\Ex{\XRV_i}=p$, the
following inequality holds for all $\varepsilon \in \Rangeab{0}{1-p}$:

\begin{align}
\Prx{\frac{1}{\NN} \sum \XRV_i \geq p + \varepsilon} & \leq \left( {\left(\frac{p}{p +
\varepsilon}\right)}^{p+\varepsilon} {\left(\frac{1 - p}{1 -p - \varepsilon}\right)}^{1 - p-
\varepsilon}\right) ^\NN \label{eq:hoeffding}
\end{align}
\end{theorem}

%%%%%%%%%%%%%%%%%%%%%%%%%%%%%%%%%%%%%%%%%%%%%%%%%%%%%%%%%%%%%%%%%%%%%%%%%%%%%%%%%%%%%%%%%%%%%%%%%%%%%%%%%%%
%%%%%%%%%%%%%%%%%%%%%%%%%%%%%%%%%%%%%%%%%%%%%%%%%%%%%%%%%%%%%%%%%%%%%%%%%%%%%%%%%%%%%%%%%%%%%%%%%%%%%%%%%%%
%%%%%%%%%%%%%%%%%%%%%%%%%%%%%%%%%%%%%%%%%%%%%%%%%%%%%%%%%%%%%%%%%%%%%%%%%%%%%%%%%%%%%%%%%%%%%%%%%%%%%%%%%%%

\phantomappendix

\begin{lemma*}[\ref{lem:single_tail}]
\label{proof:lem:single_tail}%
For any $\alpha \in \Rangeab{0}{1}$ and $j \in \PosInts$,
\begin{align*}
    \text{if $\alpha \ge 0.5$, then} \qquad&
        \Prx{\Evt{\alpha}{\{j\}}} \ge 1-\left(\Rate{\alpha}\right)^j,
            &&\text{where} \qquad \Rate{\alpha} = \frac{1}{2\alpha^\alpha (1-\alpha)^{1-\alpha}}\\
    \text{if $\alpha \le 0.5-1/j$, then} \qquad&
        \Prx{\Evt{\alpha}{\{j\}}} \le \left(\Rate{(\alpha+1/j)}\right)^{j-1}
            && \text{where $\Rate{\alpha}$ is the same as above.}
%    1
\end{align*}
\end{lemma*}
\begin{proof}
Let $\ARV_j$ be an indicator random variable which is $1$ if $j \in \A$, and $0$ otherwise.

The first inequality of the lemma follows immediately from \autoref{thm:hoeffding} by setting $\XRV_j =
\ARV_j$, $\NN=j$, $p=0.5$, and $\varepsilon=\alpha-0.5$ which yields an upper bound on
$\Prx{\EvtNot{\alpha}{\{j\}}}$ and thus a lower bound on $\Prx{\Evt{\alpha}{\{j\}}}$. Note that $\ARV_1=0$
with probability $1$, however that only decreases the probability on the left hand side of
\eqref{eq:hoeffding} so it still holds.

To prove the second inequality, we proceed as follows.
\begin{align*}
    \Prx{\Evt{\alpha}{\{j\}}} = \Prx{\frac{\SRV_j}{j} \le \alpha}
            &= \Prx{\frac{\sum_{k=1}^{j}\ARVNot_k}{j} > 1-\alpha} \\
            &= \Prx{\frac{\sum_{k=2}^{j}\ARVNot_k}{j} > 1-\alpha-\frac{1}{j}} &&\text{because $\ARVNot_1=1$ always.}\\
            &\le \Prx{\frac{\sum_{k=2}^{j}\ARVNot_k}{j-1} > 1-\alpha-\frac{1}{j}}.
\end{align*}
The second inequality of the lemma now follows immediately from \autoref{thm:hoeffding} by setting $\XRV_j =
\ARVNot_{j-1}$, $\NN=j-1$, $p=0.5$, and $\epsilon=0.5-\alpha-\frac{1}{j}$. Note that
$\Rate{1-\alpha-\frac{1}{j}}=\Rate{\alpha+\frac{1}{j}}$.
\end{proof}

\begin{theorem}[\citet*{FKG71}]
\label{thm:fkg}%
Let $\Lattice$ be a finite distributive lattice, and $\mu : \Lattice \to \PosReals$ be a function that
satisfies
\begin{align}
    \mu(x \wedge y)\mu(x\vee y) &\ge \mu(x)\mu(y), && \text{for all $x,y \in \Lattice$}.
\end{align}
Then for any two functions $f,g : \Lattice \to \PosReals$ which are either both increasing, or both
decreasing, the following inequality holds.
\begin{align}
    \left(\sum_{x\in \Lattice}f(x)g(x)\mu(x)\right)\left(\sum _{x\in \Lattice}\mu(x)\right) &\ge \left(\sum _{x\in \Lattice}f(x)\mu(x)\right)\left(\sum _{x\in \Lattice}g(x)\mu(x)\right)
\end{align}
\end{theorem}

\phantomappendix
\begin{lemma*}[\ref{lem:corr}]
\label{proof:lem:corr}%
For any $\Set,\Set' \subset \PosInts$ and $\alpha\in\RangeAB{0}{1}$, the two events $\Evt{\alpha}{\Set}$ and
$\Evt{\alpha}{\Set'}$ are positively correlated, i.e., $\Prx{\Evt{\alpha}{\Set}\cap\Evt{\alpha}{\Set'}} \ge
\Prx{\Evt{\alpha}{\Set}}\Prx{\Evt{\alpha}{\Set'}}$.
\end{lemma*}
\begin{proof}
For every $\N \in \PosInts$, define $\Set_{\N}=\Set\cap\RangeMN{1}{\N}$; similarly define $\Set'_{\N}$,
$\A_{\N}$, $\B_{\N}$, etc.

We start by proving $\Prx{\Evt{\alpha}{\Set_{\N}}\cap\Evt{\alpha}{\Set'_{\N}}} \ge
\Prx{\Evt{\alpha}{\Set_{\N}}}\Prx{\Evt{\alpha}{\Set'_{\N}}}$ for every $\N \in \PosInts$. Let $\Lattice_{\N}$
be a distributive lattice whose elements are the subsets of $\RangeMN{2}{\N}$ and whose meet/join operators
correspond to taking intersection/union. For all $\ASet\in \Lattice_\N$ let $\mu(\ASet)=1/{2^{\N-1}}$. Define
$\Evt{\alpha}{\Set_{\N}}(\ASet)$ to be an indicator function which is defined for each $\ASet \in
\Lattice_{\N}$ as
\begin{align*}
    \Evt{\alpha}{\Set_{\N}}(\ASet) &=
        \begin{cases}
            1   & \text{if $\Abs{\ASet\cap\RangeMN{1}{j}} \le \alpha j$ for all $j \in \Set_{\N}$} \\
            0   & \text{otherwise}
        \end{cases}.
\end{align*}
By invoking \autoref{thm:fkg} on lattice $\Lattice_{\N}$ and substituting $f(x)$ and $g(x)$ with
$\Evt{\alpha}{\Set_{\N}}(\ASet)$ and $\Evt{\alpha}{\Set'_{\N}}(\ASet)$ respectively we get the following
inequality.

\begin{align*}
    \left(\sum_{\ASet \subseteq \RangeMN{2}{\N}} \frac{\Evt{\alpha}{\Set_{\N}}(\ASet)\Evt{\alpha}{\Set'_{\N}}(\ASet)}{2^{\N-1}} \right)
         &\ge \left(\sum_{\ASet \subseteq \RangeMN{2}{\N}} \frac{\Evt{\alpha}{\Set_{\N}}(\ASet)}{2^{\N-1}} \right)
                \left(\sum_{\ASet \subseteq \RangeMN{2}{\N}} \frac{\Evt{\alpha}{\Set'_{\N}}(\ASet)}{2^{\N-1}} \right)
\end{align*}
Observe that the left hand side of the above inequality is exactly
$\Ex[\A]{\Evt{\alpha}{\Set_{\N}}(\A)\Evt{\alpha}{\Set'_{\N}}(\A)}=\Prx{\Evt{\alpha}{\Set_{\N}}\cap\Evt{\alpha}{\Set'_{\N}}}$
while its right hand side is exactly
$\Ex[\A]{\Evt{\alpha}{\Set_{\N}}(\A)}\Ex[\A]{\Evt{\alpha}{\Set'_{\N}}(\A)}=\Prx{\Evt{\alpha}{\Set_{\N}}}\Prx{\Evt{\alpha}{\Set'_{\N}}}$,
so we have proved that $\Prx{\Evt{\alpha}{\Set_{\N}}\cap\Evt{\alpha}{\Set'_{\N}}} \ge
\Prx{\Evt{\alpha}{\Set_{\N}}}\Prx{\Evt{\alpha}{\Set'_{\N}}}$ for every $\N \in \PosInts$.

We now prove the infinite case. For every $\N \in \PosInts$, define $\DiffSeqL_{\N}=
\Prx{\Evt{\alpha}{\Set_{\N}}\cap\Evt{\alpha}{\Set'_{\N}}}$,
$\DiffSeqR_{\N}=\Prx{\Evt{\alpha}{\Set_{\N}}}\Prx{\Evt{\alpha}{\Set'_{\N}}}$, and
$\DiffSeq_{\N}=\DiffSeqL_{\N}-\DiffSeqR_{\N}$. Observe that $\DiffSeq_{\N}$ is an infinite sequence which is
bounded in $\RangeAB{0}{1}$, so by invoking Bolzano–Weierstrass theorem we argue that it has an infinite
converging subsequence, i.e., there exists an infinite sequence of indices $\N_1 < \N_2 < \cdots$ and
$\DiffSeq^* \in \RangeAB{0}{1}$ such that $\lim_{j\to\infty} \DiffSeq_{\N_j} = \DiffSeq^*$. On the other hand
both $\DiffSeqL_{\N_j}$ and $\DiffSeqR_{\N_j}$ are decreasing sequences which are bounded below by $0$ so
they both converge, therefore
\begin{align*}
    \Prx{\Evt{\alpha}{\Set}\cap\Evt{\alpha}{\Set'}}-\Prx{\Evt{\alpha}{\Set\vphantom{\Set'}}}\Prx{\Evt{\alpha}{\Set'}}
            = \lim_{j\to\infty} \DiffSeqL_{\N_j}-\lim_{j\to\infty} \DiffSeqR_{\N_j}
            = \lim_{j\to\infty} \DiffSeq_{\N_j} = \DiffSeq^* \ge 0
\end{align*}
which proves the claim of the lemma.
% Consider the natural distributive lattice formed by , i.e., such that
%$(\A,\B) < (\A',\B')$ if{f} $\A \subset \A'$. Observe that all elements of this lattice have equal
%probability. Define $\Evt{\alpha}{\Set}(\A)$ to be an indicator function which is $1$ if
%$\Abs{\A\cap\RangeMN{1}{j}} \le \alpha j$ for all $j$, and $0$ otherwise. Obviously
%$\Prx{\Evt{\alpha}{\Set}}=\Ex[\A]{\Evt{\alpha}{\Set}(\A)}$. Observe that both $\Evt{\alpha}{\Set}(\A)$ and
%$\Evt{\alpha}{\Set'}(\A)$ are non-increasing functions over the lattice we just defined. Consequently, by FKG
%inequality~\cite{FKG71}, the two events are positively correlated, i.e.,
%$\Ex[\A]{\Evt{\alpha}{\Set}(\A)\Evt{\alpha}{\Set'}(\A)} \ge
%\Ex[\A]{\Evt{\alpha}{\Set}(\A)}\Ex[\A]{\Evt{\alpha}{\Set'}(\A)}$ which completes the proof.
\end{proof}

\end{document}